\documentstyle[12pt,epsf,epsfig]{book}

\makeatletter
\def\beq{\begin{equation}}
 \def\eeq{\end{equation}}
 \def\bea{\begin{eqnarray}}
 \def\eea{\end{eqnarray}}
\def\bq{\begin{quote}} 
\def\eq{\end{quote}}

\def\gappeq{\mathrel{\rlap {\raise.5ex\hbox{$>$}} {\lower.5ex\hbox{$\sim$}}}}

\def\lappeq{\mathrel{\rlap{\raise.5ex\hbox{$<$}} {\lower.5ex\hbox{$\sim$}}}}

\def\nothing#1{}
\newdimen\earraycolsep
\renewcommand{\thetable}{\arabic{table}}
\renewcommand{\thefigure}{\arabic{figure}}

\def\title{\chapter}
\renewcommand\chapter{\ifodd\c@page\clearpage\else\cleardoublepage\fi
		    \global\@topnum\z@
		    \@afterindenttrue
		    \secdef\@chapter\@schapter}
\def\@makechapterhead#1{%
  \vspace*{120\p@}%
  {\parindent \z@ \raggedright \reset@font
    \bfseries #1\par
    \nobreak
    \vskip 36\p@
  }}
\def\author#1 {{\pretolerance=10000 \raggedright \advance 
\leftskip by 1in \noindent #1 \vskip 1pc}}
\def\affiliation#1{{\advance\leftskip by 1in \noindent #1 \vskip -1pc}}
\def\refnote#1{{$^{\hbox{\scriptsize #1}}$}}

\def\tablenote#1 {\setbox0=\hbox{$^{\hbox{\scriptsize
#1}}$}\noindent\hangindent=\wd0 \box0}
\renewcommand\section{\@startsection{section}{1}{\z@}{2pc \@plus 1ex minus
    .2ex}{1pc \@plus .2ex}{\reset@font\normalsize\bfseries}}
\renewcommand\subsection{\@startsection{subsection}{2}{\z@}{1pc \@plus 1ex
    minus.2ex}{1pc \@plus .2ex}{\reset@font\normalsize\bfseries}}
\renewcommand\subsubsection{\@startsection{subsubsection}{3}{\parindent}
	{1pc \@plus 1ex minus.2ex}{-0.5em}{\reset@font\normalsize\bfseries}}

\def\AmS{{\protect\the\textfont2 A\kern-.1667em\lower.5ex\hbox{M}\kern-.125emS}}
\def\AmSLaTeX{\protect\AmS\-\protect\LaTeX\@}
\def\AmSTeX{\protect\AmS\-\protect\TeX\@}
\def\p@LaTeX{{\family{times}\series{m}\shape{n}
\selectfont L\kern-.36em\raise.3ex
\hbox{\scriptsize A}\kern-.15em T\kern-.1667em\lower.7ex\hbox{E}\kern-.125emX}}

\newlength{\colwidth}

\def\@oddhead{\hfil}
\def\@evenhead{\hfil}
\def\@oddfoot{{\bfseries\hfil\thepage}}
\def\@evenfoot{{\bfseries\thepage\hfil}}
\def\fnum@figure{\footnotesize\raggedright{\bfseries \figurename~\thefigure.}}
\def\fnum@table{\normalsize\raggedright{\bfseries \tablename~\thetable.}}
\long\def\@makecaption#1#2{\vskip 10\p@ {#1 #2\par}}
\long\def\@makefntext#1{\setbox0=\hbox{$\m@th^{\@thefnmark}$}\noindent
\hangindent=\wd0 \box0 #1}
\def\centerfig#1#2#3#4{\vspace*{#2}\relax\centerline {\hbox
to#1{\special{#4:#3.#4 x=#1, y=#2}\hfil}}}
\newbox\@atbox
\long\def\atable#1#2#3{\begin{table}[tbp]\centering\footnotesize
\setbox\@atbox\hbox{#2}
\parbox{\wd\@atbox}{\caption{#1}}\par\smallskip #2
\par\smallskip\parbox{\wd\@atbox}{\raggedright #3}
\end{table}}
\def\@bibitem{\noindent \hangindent=2pc \hangafter=1}
\def\thebibliography{%
\section*{REFERENCES}%
\bgroup\footnotesize
\def\newblock{\hskip .11em plus.33em minus.07em}%
\let\bibitem\@bibitem}
\def\endthebibliography{\par\egroup}
\def\@nbibitem#1{\noindent \hangindent=2pc \hangafter=1
\refstepcounter{enumi}\hbox to 2pc{\arabic{enumi}.\hfil}%
\immediate\write\@auxout{\string\bibcite{#1}{\arabic{enumi}}}}
\def\numbibliography{%
\section*{REFERENCES}%
\bgroup\footnotesize
\setcounter{enumi}{0}%
\def\newblock{\hskip .11em plus.33em minus.07em}%
\let\bibitem\@nbibitem}
\def\endnumbibliography{\par\egroup}
\makeatother

\begin{document}
\pagestyle{empty}
\begin{flushright} {CERN-TH/96-265}
\end{flushright}
\vspace*{5mm}
\begin{center} {\bf STATUS OF PRECISION TESTS OF THE STANDARD MODEL} \\
\vspace*{1cm}  {\bf G. Altarelli} \\
\vspace{0.3cm} Theoretical Physics Division, CERN \\ CH - 1211 Geneva 23 \\  and
\\ Terza Universit\`a di Roma, Rome, Italy\\
\vspace*{2cm}   {\bf CONTENT} \\ \end{center}
\vspace*{5mm}
\noindent
$$
\matrix{
 &{\rm Introduction}\hfill\cr
 &{\rm Status~of~the~Data}\hfill\cr 
 &{\rm Precision~Electroweak~Data~and~the Standard~Model}\hfill\cr &{\rm
Status~of~}~\alpha_s(m_Z)\hfill\cr &{\rm A~More~Model}-{\rm
Independent~Approach}\hfill\cr &{\rm
Conceptual~Problems~with~the~Standard~Model}\hfill\cr &{\rm
Precision~Electroweak~Tests~and~the~Search~for~New~Physics}\hfill\cr
&~~~Minimal~Supersymmetric~Standard~Model\hfill\cr &~~~Technicolor \hfill\cr
&{\rm Outlook~on~the~Search~for~New~Physics}\hfill\cr & {\rm
The~LEP2~Programme}\hfill\cr} 
$$
 
\vspace*{5cm} 

\begin{flushleft} CERN-TH/96-265\\ October 1996
\end{flushleft}
\vfill\eject

\setcounter{page}{1}
\pagestyle{plain}

\chapter{STATUS OF PRECISION TESTS OF THE STANDARD MODEL}

\author{Guido ALTARELLI}

\affiliation{Theoretical Physics Division, CERN\\
Rome, Italy}

\section{ INTRODUCTION}

	The running of LEP1 was terminated in 1995 and close-to-final results of the data
analysis are now available  and were presented at the Warsaw Conference in
July~1996\refnote{\cite{blo},\cite{ew}}. LEP and SLC started in 1989 and the first
results  from the collider run at the Tevatron were also first presented at about that
time.  I went back to my rapporteur talk at the Stanford Conference in
August~1989\refnote{\cite{sta}} and  I found the following best values quoted
there for some of the key quantities of interest for the Standard Model (SM)
 phenomenology: $m_Z$ = 91120(160) MeV; $m_t$ = 130 (50)~GeV;
$\sin^2\theta_{eff}$ = 0.23300(230) and
$\alpha_s(m_Z)$ = 0.110(10).  Now, after seven years of experimental and
theoretical work (in particular with ~16 million $Z$ events analysed
 altogether by the four LEP experiments) the corresponding numbers,
 as quoted at the Warsaw Conference, are: $m_Z$ = 91186.3(2.0) MeV; $m_t$ =
175(6) GeV;
$\sin^2\theta_{eff}$ = 0.23165(24)
 and  $\alpha_s(m_Z)$ = 0.118(3). The progress is quite evident. The top quark
has been at last found and  the errors on $m_Z$ and $\sin^2\theta_{eff}$ went
down by two and one orders of magnitude respectively. At the start  the goals of
LEP, SLC and the Tevatron were to: a)~perform precision tests of the SM at the
level of a few per mille accuracy; b)~count neutrinos ($N_{\nu}$ = 2.989(12));
c)~search for the top quark ($m_t$ = 175(6) GeV);
 d)~search for the Higgs ($m_H > 65$~GeV); e)~search for new particles (none
found). While for most of the
 issues the results can be summarized in very few bits, as just shown,  it is by
far more complex for the first one.  The validity of the SM has been confirmed to
a level that I can say was unexpected at the
 beginning. This is even more true after Warsaw. Contrary to the situation
presented at  the winter '96 Conferences we are now left with no significant
evidence for departures  from the SM. The discrepancy on $R_c$ has completely
disappeared, that on
$R_b$ has been much reduced, and so on, and no convincing hint of new physics is
left in the data (also including the first  results from LEP2). The impressive
success of the SM poses strong limitations on the possible forms of  new physics.
Favoured are models of the Higgs sector and of new physics that preserve the SM
structure  and only very delicately improve it, as is the case for fundamental
Higgs(es) and Supersymmetry.
 Disfavoured are models with a nearby strong non-perturbative regime that  almost
inevitably would affect the radiative corrections, as for composite Higgs(es) or
for technicolor and its variants. 

\section{STATUS OF THE DATA}

	The relevant new electroweak data together with their SM values are presented in
table 1.  The SM values correspond to a fit in terms of $m_t$, $m_H$ and
$\alpha_s(m_Z)$, described later in sect.~3, eq.~(\ref{9}),  of all the available
data including the CDF/D0 value of $m_t$ . A number of comments on the novel
aspects of the data are now in order.

\vspace*{0.5cm}
\begin{center} Table 1\\
\vglue.3cm
\begin{tabular}{|l|l|l|l|}
\hline Quantity&Data (Warsaw '96)	& Standard Model & Pull\\
\hline
$m_Z$ (GeV)&91.1863(20)	&91.1861 &~~0.1\\
$\Gamma_Z$ (GeV)	&2.4946(27) & 2.4960 & $-0.5 $\\
$\sigma_h$ (nb)	&41.508(56)	&41.465 & ~~0.8\\
$R_h$	&20.788(29)	&20.757 & ~~0.7\\ 
$R_b$ &0.2178(11)	&0.2158 & ~~1.8\\ 
$R_c$&	0.1715(56)&	0.1723 & $-0.1$ \\
$A^l_{FB}$&  0.0174(10)	&0.0159 & ~~1.4 \\
$A_\tau$ &	0.1401(67)	&0.1458 & $-0.9$ \\
$A_e$	&0.1382(76) &0.1458& $-1.0$\\ 
$A^b_{FB}$ & 	0.0979(23) &0.1022 & $-1.8$ \\
$A^c_{FB}$& 	0.0733(48)	&0.0730 & ~~0.1\\ 
$A_b$	&SLD direct 0.863(49) &0.935 & $-2.2$\\ &LEP indir. 0.895(23) &&\\ &Average
0.889(21)	  &&\\
$A_c$	& SLD direct 0.625(84) &0.667 & $-0.2$\\ &LEP indir. 0.670(44) && \\
&Average 0.660(39)	&& \\
$\sin^2\theta_{eff}({\rm\hbox{LEP-combined}})$ & 0.23200(27) & 0.23167 & ~~1.2\\
$A_{LR}\rightarrow  \sin^2\theta_{eff}$& 0.23061(47) &	0.23167	& $-2.2$ \\
$m_W$ (GeV) & 80.356(125)	&80.353& ~~0.3\\
$m_t$ (GeV)	&175(6)	&172 & ~~0.5\\
\hline
\end{tabular}
\end{center}
\vglue.3cm What happened to $R_c$? The tagging method for charm is based on the
reconstruction of exclusive final channels. This is rather complicated and
depends on branching ratios and on the probability that a charm quark fragments
into given hadrons. A shift in the measured value of the branching ratio for
$D^0\rightarrow K^-\pi^+$ and the measurement at LEP of $P(c\rightarrow D^*$),
acting on
$R_c$  in the same direction, have been sufficient to restore a perfect agreement
with the SM. 

	What happened to $R_b$? The old result at the winter '96 Conferences was
(assuming the SM value for
$R_c$) $R_b$ = 0.2202(16). The present official average, shown in table 1, is
much lower and only
$1.8\sigma$ away from the SM value. The essential difference is the result of a
new-from-scratch, much improved analysis from ALEPH, which is given by
\refnote{\cite{blo},\cite{ew}}
 \beq
 R_b = 0.2161\pm0.0014~~~~~~~~~~~~~(\rm{ALEPH})~.
\label{1}
 \eeq
 In fact if one combines the average of the  ``old" measurements, given above,
with the ``new" ALEPH result one practically finds the official average given by
the electroweak LEP working group and reported in table 1. This happens to be
true in spite of the fact that in the correct procedure one has to take away the
ALEPH contribution, now superseded, from the ``old" average and add to it some
newly presented refinements to some of the ``old" analyses. In view of this, it
is clear that the change is mainly due to the new ALEPH result. There are
objective improvements in this new analysis. Five mutually exclusive tags are
simultaneously used in order to decrease the sensitivity to individual sources of
systematic error. Separate vertices are reconstructed in the two hemispheres of
each event to minimize correlations between  the hemispheres. The implementation
of a mass tag on the tracks from each vertex reduces the charm background that
dominates the systematics. As a consequence it appears to me that the weight of
the new analysis in the combined value should be larger than what is obtained
from the stated errors. In view of the ALEPH result the necessity of new physics
in $R_b$ has disappeared, while the possibility of some small deviation (more
realistic than before) of course is still there. In view of the importance of
this issue the other collaborations will go back to their data and freshly
reconsider their analyses with the new improvements taken into account.  

	It is often stated that there is a $3\sigma$ deviation on the measured value of
$A_b$ with respect to the SM expectation\refnote{\cite{blo},\cite{ew}}. But in
fact that  depends on how the data are combined. In my opinion  one should rather
talk of a
$2\sigma$ effect. Let us discuss this point in detail. $A_b$ can be measured
directly at SLC, taking advantage of the beam longitudinal polarization. SLD finds
 \beq
 A_b = 0.863\pm0.049~~~~~~~~~~~~~(\rm{SLD~direct:-1.5\sigma})~,
 \label{2}
 \eeq
 where the  discrepancy with respect to the SM value, $A^{SM}_b$    = 0.935, has
also been indicated. At LEP one measures
$A^{FB}_b$    = 3/4 $A_eA_b$. As seen in table 1, the value found is somewhat
below the SM prediction. One can  then derive $A_b$ by using the value of $A_e$
obtained, using lepton universality, from the measurements of $A^{FB}_l$,
$A_\tau$, $A_e$: $A_e$ = 0.1466(33):
 \beq
 A_b = 0.890\pm0.029~~~~~(\rm{LEP}, ~A_e~ {\rm from~LEP:} -1.6\sigma)~.
 \label{3}
 \eeq
 By combining the two above values one obtains
 \beq
 A_b = 0.883\pm0.025~~~~~(\rm{LEP+SLD}, A_e ~{\rm from~ LEP}: -2.1\sigma)~.
 \label{4}
 \eeq The LEP electroweak working group combines the SLD result with the LEP
value for $A_b$ modified by adopting for
$A_e$ the SLD+LEP average value, which also includes $A_{LR}$ from SLD, $A_e$ =
0.1500(25):
	\beq
 A_b = 0.867\pm0.020~~~~~(\rm{LEP+SLD},~A_e ~{\rm from~ LEP+SLD:} -3.1\sigma)~.
 \label{5}
 \eeq
	 There is nothing wrong with that but, in this case, the well-known $\sim
2\sigma$ discrepancy of $A_{LR}$ with respect to $A_e$ measured at LEP and also
to the SM, which is not related to the $b$ couplings, further contributes to
inflate the number of
$\sigma$'s. Since the $b$ couplings are more suspect than the lepton couplings it
is perhaps wiser to obtain $A_b$ from LEP by using the SM value for
$A_e$:  $A^{SM}_e$   = 0.1458(16), which gives
	\beq
 A_b = 0.895\pm0.023~~~~~(\rm{LEP},~A_e=A^{SM}_e: -1.7\sigma)~.
 \label{6}
 \eeq
	 Finally, combining the last value with SLD we have
 \beq
 A_b = 0.889\pm0.021~~~~~(\rm{LEP+SLD},~A_e=A^{SM}_e: -2.2\sigma)~.
 \label{7} 
\eeq	  Note that these are the values reported in table 1.

	Finally if one looks at the values of $\sin^2\theta_{eff}$ obtained from
different observables, shown in fig.~1 , one notices that the value obtained from
$A^{FB}_l$    is somewhat low (indeed quite in agreement with its determination
by SLD from
$A_{LR}$).  Looking closer, this is due to the FB asymmetry of the $\tau$ lepton
that, systematically in all four LEP experiments, has a central value above that
of $e$ and
$\mu$ \refnote{\cite{blo},\cite{ew}}. The combined value for the  $\tau$ channel
is
$A^{FB}_\tau$  =0.0201(18) while the combined average of $e$ and $\mu$ is
$A^{FB}_{e/\mu}$    = 0.0162(11). On the other hand $A_\tau$ and  $\Gamma_\tau$
appear normal. In principle these two facts are not incompatible, because the FB
lepton asymmetries are very small. The extraction of
$A^{FB}_\tau$   from the data on the angular distribution of $\tau$'s could be
biased if the imaginary part of the continuum was altered by some non-universal
new physics effect\refnote{\cite{car}}. But a more trivial experimental problem
is at the moment more plausible.

\begin{figure}
\hglue3.5cm
\epsfig{figure=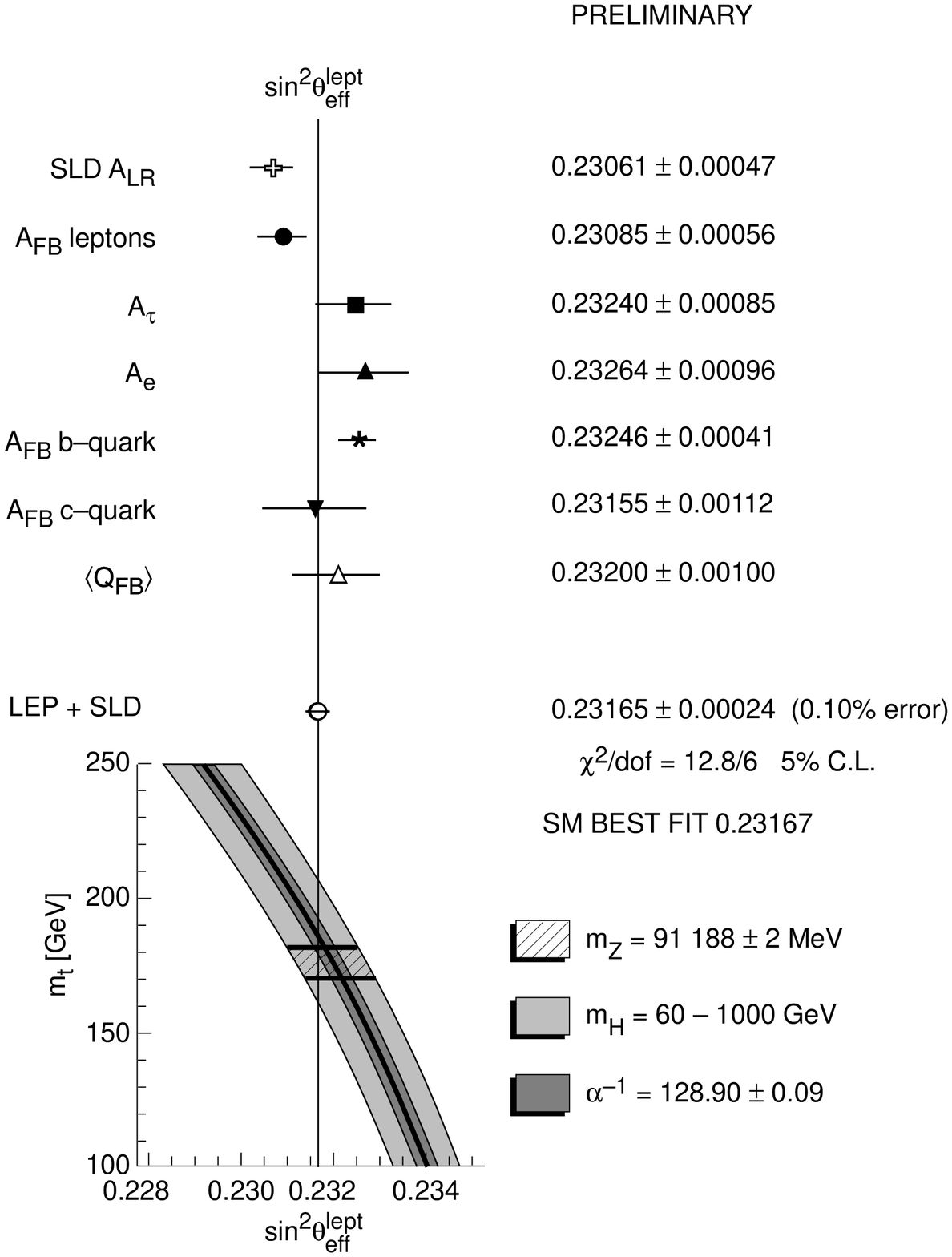,width=8.0cm}
\begin{center} Figure 1 \end{center}
\end{figure}

	The distribution of measured values of $\sin^2\theta_{eff}$, as it is summarized
in fig.~1, is somewhat wide ($\chi^2$/d.o.f. = 2.13) with $A^{FB}_l$    and
$A_{LR}$  far on one side and $A^{FB}_b$      on the other side. In view of this
it would perhaps be appropriate to enlarge the error on the average from 
$\pm$0.00024 up to  
$\pm\sqrt{2.13}~0.00024$ =  $\pm$0.00034, according to the recipe adopted by the
Particle Data Group. Thus from time to time in the following we will use the
average
\beq
\sin^2\theta_{eff}=0.23165\pm0.00034
 \label{8}
 \eeq	

\section{PRECISION ELECTROWEAK DATA AND THE STANDARD MODEL}
 
	For the analysis of electroweak data in the SM one starts from the input
parameters: some of them,
$\alpha$, $G_F$ and $m_Z$, are very well measured, some other, $m_{f_{light}}$,
$m_t$ and
$\alpha_s(m_Z)$  are only approximately determined, while $m_H$ is largely
unknown. With respect to
$m_t$ the situation has much improved since the CDF/D0 direct measurement of the
top quark mass\refnote{\cite{tip}}. From the input parameters one computes the
radiative corrections\refnote{\cite{10},\cite{11}} to a sufficient precision to
match the experimental capabilities. Then one compares the theoretical
predictions and the data for the numerous observables that have been measured,
checks the consistency of the theory and derives constraints on $m_t$,
$\alpha_s(m_Z)$, and hopefully also on $m_H$. 

	Some comments on the least known of the input parameters are now in order.  The
only practically relevant terms where precise values of the light quark masses,
$m_{f_{light}}$, are needed are those related to the hadronic contribution to the
photon vacuum polarization diagrams that determine
$\alpha(m_Z)$. This correction is of order 6$\%$, much larger than the accuracy
of a few per mille of the precision tests. Fortunately, one can use the actual
data to in principle solve the related ambiguity. But we shall see that the
left-over uncertainty is still one of the main sources of theoretical error. As
is well known\refnote{\cite{piet}--\cite{21}}, the QED running coupling is given
by:
\begin{eqnarray}
\alpha(s) &=& \frac{\alpha}{1-\Delta \alpha(s)}\nonumber \\
\Delta \alpha(s) &= &\Pi(s) = \Pi_\gamma(0) - {\rm Re} \Pi_\gamma(s)~,
\label{2a}
\end{eqnarray} where $\Pi(s)$ is proportional to the sum of all 1-particle
irreducible vacuum polarization diagrams. In perturbation theory
$\Delta\alpha(s)$ is given by
\begin{equation}
\Delta \alpha(s) = \frac{\alpha}{3\pi} \sum_f Q^2_f N_{Cf}\left( \log
\frac{2}{m^2_f} - \frac{5}{3} \right)~,
\label{3a}
\end{equation} where $N_{Cf} = 3$ for quarks and 1 for leptons. However, the
perturbative formula is only reliable for leptons, not for quarks (because of the
unknown values of the effective quark masses). Separating the leptonic, the light
quark and the top quark contributions to $\Delta\alpha(s)$ we have:
\begin{equation}
\Delta\alpha(s) = \Delta\alpha(s)_1 + \Delta\alpha(s)_h + \Delta\alpha(s)_t
\label{4a}
\end{equation}		 with\refnote{\cite{21}}:
\begin{equation}
\Delta\alpha(s)_1 = 0.0331421~;~~\Delta\alpha(s)_t =
\frac{\alpha}{3\pi}~\frac{4}{15}~\frac{m^2_Z}{m^2_t} = -0.000061~.
\label{5a}
\end{equation} Note that in QED there is decoupling so that the top quark
contribution approaches zero in the large $m_t$ limit. For $\Delta\alpha(s)_h$
one can use (\ref{2a}) and the Cauchy theorem to obtain the representation:
\begin{equation}
\Delta\alpha(m^2_Z)_h = -\frac{\alpha m^2_Z}{3\pi}{\rm Re}
\int^\infty_{4m^2_\pi}\frac{ds}{s}~\frac{R(s)}{s-m^2_Z-i\epsilon}
\label{6a}
\end{equation} where $R(s)$ is the familiar ratio of the hadronic to the
point-like $\ell^+\ell^-$ cross-section from photon exchange in $e^+e^-$
annihilation. At $s$ large, one can use the perturbative expansion for $R(s)$
while at small $s$ one can use the actual data. 

	Recently there has been a lot of activity on this subject and a number of
independent new estimates of $\alpha(m_Z)$  have appeared in the
literature\refnote{\cite{piet}}. In table~2 we report the results of these new
computations together with the most significant earlier determinations
(previously the generally accepted value was that of Jegerlehner in
1991\refnote{\cite{15}}).

\begin{center} Table 2\\
\vglue.3cm
\begin{tabular}{|l|c|l|l|}
\hline ~~~~Author & Year and Ref.	& ~~~~$\Delta\alpha(m^2_Z)_h$ &
~~~$\alpha(m^2_Z)^{-1}$\\
\hline			 Jegerlehner & 1986 \cite{12} & 0.0285~~$\pm$ 0.0007 & 128.83 $\pm$
0.09\\ Lynn et al. & 1987 \cite{13} & 0.0283~~$\pm$ 0.0012 & 128.86$\pm$0.16\\
Burkhardt et al. & 1989 \cite{14} & 0.0287~~$\pm$ 0.0009 & 128.80 $\pm$ 0.12\\
Jegerlehner & 1991 \cite{15} & 0.0282~~$\pm$ 0.0009 & 128.87$\pm$0.12\\ Swartz &
1994 \cite{16} & 0.02666$\pm$ 0.00075 & 129.08$\pm$0.10\\ Swartz (rev.) & 1995
\cite{17} & 0.0276~~$\pm$ 0.0004 & 128.96 $\pm$ 0.06\\ Martin et al. & 1994
\cite{18} & 0.02732$\pm$ 0.00042 & 128.99 $\pm$ 0.06\\ Nevzorov et al. & 1994
\cite{19} & 0.0280~~$\pm$ 0.0004 & 128.90 $\pm$ 0.06\\ Burkhardt et al. & 1995
\cite{20} & 0.0280~~$\pm$ 0.0007 & 128.89 $\pm$ 0.09\\ Eidelman et al. & 1995
\cite{21} & 0.0280~~$\pm$ 0.0007 & 128.90 $\pm$ 0.09\\
\hline
\end{tabular}
\end{center}
\vspace*{0.5cm}

The differences among the  recent determinations are due to the procedures
adopted for fitting the data and treating the errors, for performing the
numerical integration, etc. The differences are also due to the threshold chosen
to start the application of perturbative QCD at large $s$ and to the value
adopted for $\alpha_s(m_Z)$. For example, in its first version
Swartz\refnote{\cite{16}} used parametric forms to fit the data, while most of
the other determinations use a trapezoidal rule to integrate across the data
points. It was observed that the parametric fitting introduces a definite
bias\refnote{\cite{17}}. In fact Swartz gets systematically lower results for all
ranges of $s$. In its revised version\refnote{\cite{17}} Swartz improves his
numerical procedure. Martin et al.\refnote{\cite{18}} use perturbative QCD down to
$\sqrt{s}$    = 3 GeV (except in the upsilon region) with 
$\alpha_s(m_Z)$ = 0.118$\pm$0.007. Eidelman et al.\refnote{\cite{21}} only use
perturbative QCD for $\sqrt{s}> 40$ GeV and with $\alpha_s(m_Z)$ = 0.126$\pm$
0.005, i.e. the value found at LEP. They use the trapezoidal rule. Nevzorov et
al.\refnote{\cite{19}} make a rather crude model with one resonance per channel
plus perturbative QCD with
$\alpha_s(m_Z)$ = 0.125 $\pm$ 0.005. Burkhardt et al.\refnote{\cite{20}} use
perturbative QCD for
$\sqrt{s}> 12$~GeV, but with a very conservative error on
$\alpha_s(m_Z)$ = 0.124 $\pm$ 0.021. This value was determined
\refnote{\cite{22}} from
$e^+e^-$ data below LEP energies. The excitement produced by the original claim by
Swartz\refnote{\cite{16}} of a relatively large discrepancy with respect to the
value obtained by Jegerlehner\refnote{\cite{15}} resulted in a useful debate. As
a conclusion of this re-evaluation of the problem the method of Jegerlehner has
proved its solidity. As a consequence I think that the recent update by Eidelman
and Jegerlehner\refnote{\cite{21}} gives a quite reliable result (which is the
one used by the LEP groups and in the following). Also, I do not think that a
smaller error than quoted by these authors can be justified.

	As for the strong coupling $\alpha_s(m_Z)$ we will discuss in detail the
interesting recent developments in sect. 4. The world average central value is
quite stable around 0.118, before and after the most recent results. The error is
going down because the dispersion among the different measurements is much
smaller in the most recent set of data. The error is taken to be between
$\pm$0.003 and
$\pm$0.005, depending on how conservative one wants to be. Thus in the following
our reference value will be $\alpha_s(m_Z)$ = 0.118 $\pm$ 0.005.

	Finally a few words on the current status of the direct measurement of $m_t$. The
error is rapidly going down. It was $\pm$9 GeV before the Warsaw Conference, it
is now
$\pm$6 GeV
\refnote{\cite{tip}}. I think one is soon approaching a level where a more careful
investigation of the effects of colour rearrangement on the determination of
$m_t$ is needed. One wants to determine the top quark mass, defined as the
invariant mass of its decay products (i.e. $b+W+$ gluons + $\gamma$'s). However,
due to the need of colour rearrangement, the top quark and its decay products
cannot be really isolated from the rest of the event. Some smearing of the mass
distribution is induced by this colour crosstalk, which involves the decay
products of the top, those of the antitop and also the fragments of the incoming
(anti)protons. A reliable quantitative computation of the smearing effect on the
$m_t$  determination is difficult because of the importance of non-perturbative
effects. An induced error of the order of a few GeV on $m_t$ is reasonably
expected. Thus further progress on the $m_t$ determination demands tackling this
problem in more depth. 

	The measured top production cross section is in fair agreement with the QCD
prediction, but the central value is a bit large (see
fig.~2)\refnote{\cite{wil}}. The world average for the cross section times
branching ratio is $\sigma B= 6.4\pm1.3$ pb and the QCD prediction for
$\sigma$ is $\sigma_{QCD}$ = 4.75 $\pm$ 0.65 pb \refnote{\cite{sigtop}}. Thus the
branching ratio $B = B(t\rightarrow bW$) cannot be far from 100\%   unless there
is also some additional production mechanism from new physics.

\begin{figure}
\hglue3.5cm
\epsfig{figure=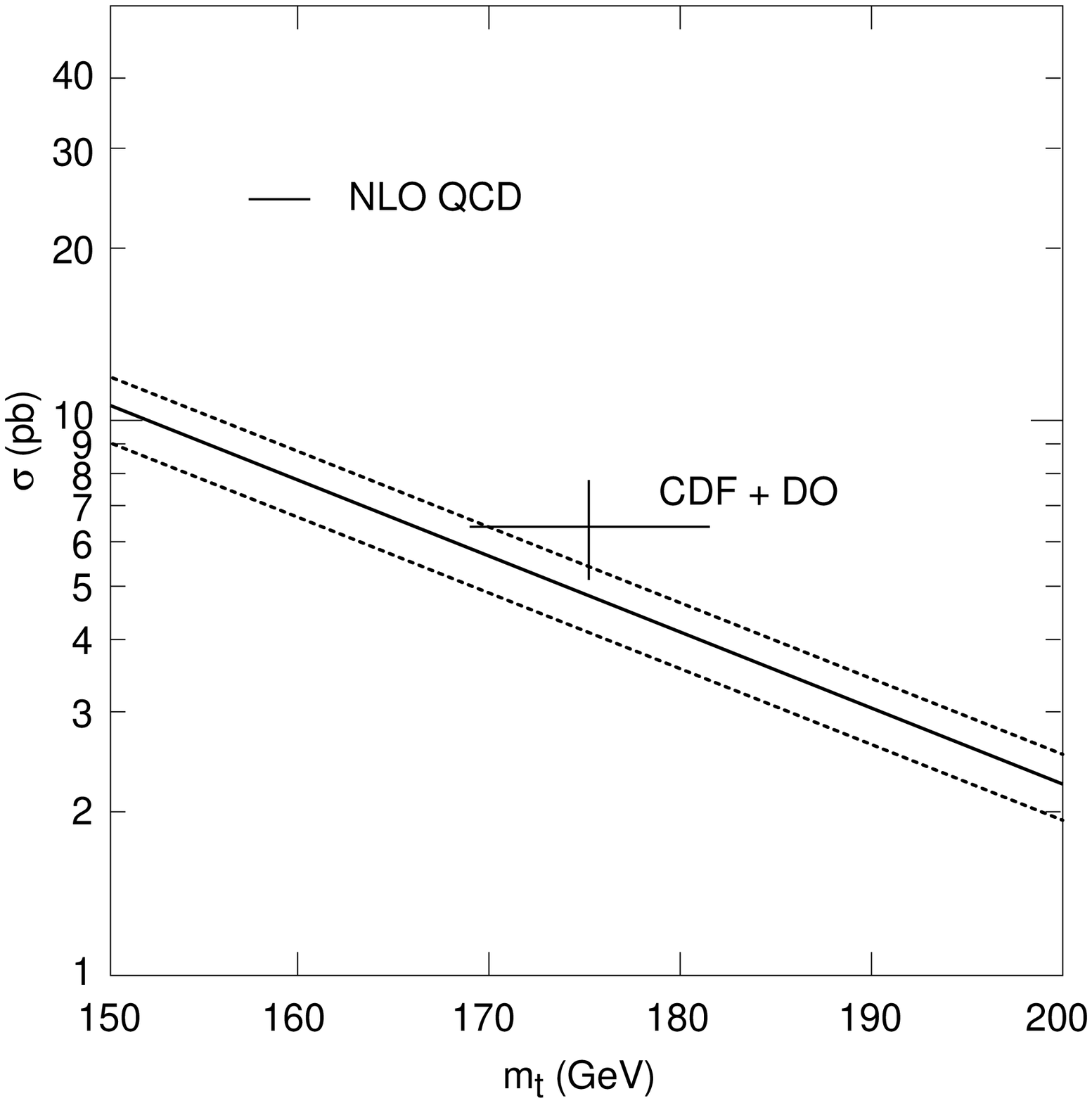,width=8.5cm}
\begin{center} Figure 2 \end{center}
\end{figure}

	In order to appreciate the relative importance of the different sources of
theoretical errors for precision tests of the SM, I report in table 3  a
comparison for the most relevant observables, evaluated using
ref.~\refnote{\cite{23}}.

\begin{table} Table 3: Errors from different sources: $\Delta^{exp}_{now}$    is
the present experimental error;  
$\Delta\alpha^{-1}$ is the impact of $\Delta\alpha^{-1}=\pm0.09$;  $\Delta_{th}$ 
is the estimated theoretical error from higher orders; $\Delta m_t$ is from
$\Delta m_t =\pm 6 $GeV;
$\Delta m_H$ is from $\Delta m_H$ = 60--1000 GeV; $\Delta\alpha_s$ corresponds to
$\Delta\alpha_s=\pm0.005$. The epsilon parameters are defined in
ref.~\refnote{\cite{24}}.

\begin{center}
\begin{tabular}{|l|l|l|l|l|l|l|}
\hline Parameter& $\Delta^{exp}_{now}$ & $\Delta \alpha^{-1}$ & $\Delta_{th}$ &
$\Delta m_t$ & $\Delta m_H$ & $\Delta \alpha_s$ \\
\hline
$\Gamma_Z$ (MeV) & $\pm$2.7 & $\pm$0.7 & $\pm$0.8 & $\pm$1.4 & $\pm$4.6 &
$\pm$2.7 \\
$\sigma_h$ (pb) & 56 & 1 & 4.3 & 3.3 & 4 & 2.7\\
$R_h \cdot 10^3$ & 29 & 4.3 & 5 & 2 & 13.5 & 34 \\
$\Gamma_l$ (keV) & 110 & 11 & 15 & 55 & 120 & 6\\
$A^l_{FB}\cdot 10^4$ & 10 & 4.2 & 1.3 & 3.3 & 13 & 0.3 \\
$\sin^2\theta\cdot 10^4$ & $\sim$3 & 2.3 & 0.8 & 1.9 & 7.5 & 0.15\\
$m_W$~(MeV) & 125 & 12 & 9 & 37 & 100& 4 \\
$R_b \cdot 10^4$ & 11 & 0.1 & 1 & 2.1 & 0.25 & 0\\
$\epsilon_1\cdot 10^3$ & 1.3 & & $\sim$0.1 & & & 0.4\\
$\epsilon_3\cdot 10^3$ & 1.4 & 0.6 & $\sim$0.1 & & & 0.25\\
$\epsilon_b\cdot 10^3$ & 3.2 & & $\sim$0.1 & & & 2\\
\hline
\end{tabular}
\end{center}
\end{table}
	What it is  important to stress is that the ambiguity from $m_t$, once by far the
largest one, is by now smaller than the error from $m_H$. We also see from table 3
that the error from $\Delta\alpha(m_Z)$ is expecially important for
$\sin^2\theta_{eff}$  and, to a lesser extent, is also sizeable for
$\Gamma_Z$ and $\epsilon_3$.  

	We now discuss fitting the data in the SM. As the mass of the top quark is now
rather precisely known from CDF and D0 one must distinguish between two different
types of fits. In one type one wants to answer the question: Is $m_t$  from
radiative corrections in agreement with the direct measurement at the Tevatron?
For answering this interesting but somewhat limited question, clearly one must
exclude the CDF/D0 measurement of
$m_t$ from the input set of data. Fitting the data in terms of $m_t$, $m_H$ and
$\alpha_s(m_Z)$ one finds the results shown in table 4\refnote{\cite{ew}}. 

\begin{center} Table 4\\
\vglue.3cm
\begin{tabular}{|l|l|l|l|}
\hline Parameter & LEP & LEP + SLD & All $\not= m_t$\\
\hline
$\alpha_s(m_Z)$ & 0.1211(32) & 0.1200(32) & 0.1202(33) \\
$m_t$ (GeV) & 155(14) & 156(11) & 157(10)\\
$m_H$ (GeV) & 86($+202-14$) & 48($+83-26$) & 149($+148-82$)\\
$(m_H)_{\rm MAX}$ at 1.64$\sigma$ & 417 & 184 & 392\\
$\chi^2/dof$ & 5/8 & 18/11 & 18/13\\
\hline
\end{tabular}
\end{center}
\vglue.3cm					

The extracted value of $m_t$ is typically a bit too low. For example, from LEP
data alone one finds
$m_t$ = 155(14) GeV. But this is simply due to $R_b$ being taken from the official
average:
$R_b$ = 0.2178(11). If $m_H$ is not fixed the fit prefers lower values of $m_t$ to
adjust $R_b$. In fact by removing $R_b$  from the input data one increases the
central value of $m_t$ from 155 to 171 GeV. In this context it is important to
remark that fixing $m_H$ at 300 GeV, as is often done,  is by now completely
obsolete, because it introduces a strong bias on the fitted value of $m_t$. The
change induced on the fitted value of $m_t$ when moving $m_H$ from 300 to 65 or
1000 GeV is in fact larger than the error on the direct measurement of $m_t$. 

	In a more general type of fit, e.g. for determining the overall consistency of
the SM or the best present estimate for some quantity, say $m_W$, one should of
course not ignore the existing direct determination of $m_t$. Then, from all the
available data, including
$m_t$ = 175(6) GeV, by fitting
$m_t$, $m_H$ and $\alpha_s(m_Z)$ one finds (with $\chi^2$/d.o.f. = 19/14)
\refnote{\cite{ew}} (see also\refnote{\cite{gur}}): 
\bea
	m_t &=& 172\pm6~{\rm GeV}~,\nonumber \\
			m_H &=& 149+148-82~{\rm (or}~m_H < 392~{\rm GeV~at}~1.64\sigma) \nonumber\\
\alpha_s(m_Z) &=& 0.1202 \pm 0.0033~.
 \label{9}
 \eea
 This is the fit reported in table~1. The corresponding fitted values of
$\sin^2\theta_{eff}$ and $m_W$ are:
 \bea
		\sin^2\theta_{eff}& = & 0.23167\pm0.0002			\nonumber \\
			m_W &=& 80.352\pm0.034~{\rm GeV}~.
\label{10}
 \eea  The error of 34 MeV on $m_W$  clearly sets up a goal for the direct
measurement of $m_W$ at LEP2 and the Tevatron.

\section{STATUS OF $\alpha_s(m_Z)$}

	There are important developments in the experimental determination of
$\alpha_s(m_Z)$\refnote{\cite{sch}}. There is now a much better agreement between
the different methods of measuring $\alpha_s(m_Z)$. In fact the value of
$\alpha_s(m_Z)$ from the $Z$ line shape went down and the values from scaling
violations in deep inelastic scattering and from lattice QCD went up. We will
discuss these developments in detail in the following.

	The value of $\alpha_s(m_Z)$ from the $Z$ line shape (assuming that the SM is
valid for $\Gamma_h$, which is not completely evident in view of $R_b$) went down
for two reasons\refnote{\cite{blo},\cite{ew}}. First the value extracted from
$R_h$ only, which was
$\alpha_s(m_Z)$ = 0.126(5), is now down to
$\alpha_s(m_Z)$ = 0.124(5). Second the value from all the $Z$ data changed from
$\alpha_s(m_Z)$ = 0.124(5) down to $\alpha_s(m_Z)$ = 0.120(4), which corresponds
to the fit in eq.~(\ref{9}) . The main reason for this decrease is the new value
of $\sigma_h$ (with a sizeably smaller error than in the past) that prefers a
smaller
$\alpha_s(m_Z)$. However this determination depends on the assumption that
$\Gamma_b$ is given by the SM. We recall that $R_b$ itself with good
approximation is independent of $\alpha_s$, but its deviation from the SM would
indicate an anomaly in $\Gamma_b$ hence in
$\Gamma_h$. Taking a possible anomaly in $R_b$ into account the $Z$ line shape
determination of
$\alpha_s(m_Z)$ becomes approximately:
 \beq
			\alpha_s(m_Z)= (0.120\pm0.004) - 4 \delta R_b~.				 
\label{11} 
\eeq
 If the ALEPH value for $R_b$  (see eq~.(\ref{1})) is adopted, the central value
of $\alpha_s(m_Z)$ is not much changed, but of course the error on $\delta R_b$
is transferred on 
$\alpha_s(m_Z)$, which becomes
		\beq
			\alpha_s(m_Z)= 0.119\pm0.007~.
\label{12} 
\eeq	  If, instead, one takes $R_b$ from table 1 one obtains a much smaller
central value: 
\beq
			\alpha_s(m_Z)= 0.112\pm0.006	 
\label{13} 
\eeq
			
Summarizing: the $Z$ line shape result for $\alpha_s(m_Z)$, obtained with the
assumption that $\Gamma_h$ is given by the SM, went down a bit. The central value
could be shifted further down if $R_b$ is in excess with respect to the SM. 

	While $\alpha_s(m_Z)$ from LEP goes down, $\alpha_s(m_Z)$ from the scaling
violations in deep inelastic scattering goes up. To me the most surprising result
from Warsaw was the announcement by the CCFR collaboration that their well-known
published analysis of
$\alpha_s(m_Z)$ from $xF_3$ and $F_2$ in neutrino scattering off Fe target  is now
superseded by a reanalysis of the data based on better energy
calibration\refnote{\cite{har}}. We recall that their previous result,
$\alpha_s(m_Z)$ =
 0.111(3~exp), being in perfect agreement with the value obtained from e/$\mu$
beam data by BCDMS and SLAC combined\refnote{\cite{virc}},  $\alpha_s(m_Z)$ =
0.113(3~exp), convinced most of us that the average value of $\alpha_s(m_Z)$ from
deep inelastic scattering was close to 0.112. Now the new  result presented in
Warsaw is
\refnote{\cite{har}, \cite{sch}}
\beq
\alpha_s(m_Z)= 0.119\pm0.0015 {\rm (stat)}\pm0.0035 {\rm (syst)} \pm0.004
(th)~~~~(\rm {CCFR-revised)}~,	
\label{14} 
\eeq  where the error also includes the collaboration estimate of the theoretical
error from scale and renormalization scheme ambiguities. As a consequence the new
combined value of
$\alpha_s(m_Z)$ from scaling violations in  deep inelastic scattering is given by
  \beq
			\alpha_s(m_Z)= 0.115\pm0.006~,
	 \label{15}
 \eeq with my more conservative estimate, of the common theoretical error
(Schmelling, the rapporteur in Warsaw quotes $\pm  0.005$
\refnote{\cite{sch}}). If we compare eq.~(\ref{15}) with LEP eq.~(\ref{9}), we see
that, whatever our choice of theoretical errors is,  there is no need for any new
physics in $R_b$ to fill the gap between the two determinations of
$\alpha_s(m_Z)$.

	Finally $\alpha_s(m_Z)$ from lattice QCD is also going up \refnote{\cite{fly}}.
The main new development is a theoretical study of the error associated with the
extrapolation from unphysical values of the light quark masses, which is used in
the lattice extraction of $\alpha_s(m_Z)$ from quarkonium splittings. According to
ref.~\refnote{\cite{gri}} this effect amounts to a shift upward of +0.003 in the
value of
$\alpha_s(m_Z)$. From the latest unquenched determinations of $\alpha_s(m_Z)$,
Flynn, the rapporteur in Warsaw\refnote{\cite{fly}}, gives an average of 
0.117(3). But the lattice measurements of 
$\alpha_s(m_Z)$ moved very fast over the last few years. At the Dallas conference
in 1992, the quoted value (from quenched computations) was $\alpha_s(m_Z)$ =
0.105(4)
\refnote{\cite{dal}}, while at Beijing in 1995 the claimed value was
$\alpha_s(m_Z)$ = 0.113(2) but the error was estimated to be $\pm   0.007$ by the
rapporteur Michael\refnote{\cite{mich}}. So, with the present central value, I
will keep this more conservative error in the following: 
\beq
			\alpha_s(m_Z)= 0.117\pm0.007~.	 
\label{16} 
\eeq
				
	To my knowledge, there are no other important new results on the determination of
$\alpha_s(m_Z)$. Adding a few more well-established measurements of
$\alpha_s(m_Z)$ we have table~5, where the errors denote my personal view of the
weights  the different methods should have in the average (in brackets Th and Exp
are labels that indicate whether the dominant error is theoretical or
experimental).
\vglue.3cm
\begin{table}
	\begin{center} Table 5\\
\vglue.3cm
\begin{tabular}{|l|ll|}
\hline Measurements & \multicolumn{2}{c|}{$\alpha_s(m_Z)$}\\ 
\hline
$R_{\tau}$ & 0.122 $\pm$ 0.007 & (Th)\\ Deep Inelastic Scattering & 0.115 $\pm$
0.006 & (Th)\\
$Y_{\rm decay}$ & 0.112 $\pm$ 0.010 & (Th)\\ Lattice QCD & 0.117 $\pm$ 0.007 &
(Th)\\
$Re^+e^-(\sqrt s < 62~{\rm GeV}$) & 0.124 $\pm$ 0.021 & (Exp)\\ Fragmentation
functions in $e^+e^-$ & 0.124 $\pm$ 0.010 & (Th)\\ Jets in $e^+e^-$ at and below
the $Z$ & 0.121 $\pm$ 0.008 & (Th)\\
$Z$ line shape (taking $R_b$ from ALEPH) & 0.119 $\pm$ 0.007 & (Exp)\\
\hline
\end{tabular}
\end{center}
\end{table}
\vglue.3cm
			  The average value given as 
\beq
			\alpha_s(m_Z)= 0.118\pm0.003	 
\label{17} 
\eeq
 is very stable. The same value was quoted by Schmelling, a rapporteur at the
Warsaw Conference\refnote{\cite{sch}}, with a different treatment of errors. Had
we used
$\alpha_s(m_Z)$ from the $Z$ line shape assuming the SM value for $R_b$, i.e.
$\alpha_s(m_Z)$ = 0.120$\pm$0.004, the average value would have been 0.119. To be
safe one could increase the error to $\pm0.005$.

\section{A MORE MODEL-INDEPENDENT APPROACH}

	We now discuss an update of the epsilon analysis\refnote{\cite{24}}. The epsilon
method is more complete and less model-dependent than the similar approach based
on the variables $S, T$ and $U$ \refnote{\cite{26}--\cite{29}} which, from the
start, necessarily assumes dominance of vacuum polarization diagrams from new
physics and truncation of the $q^2$ expansion of the corresponding amplitudes. In
a completely model-independent way we define\refnote{\cite{24}} four variables,
called
$\epsilon_1$, $\epsilon_2$,
$\epsilon_3$ and
$\epsilon_b$, that are precisely measured and can be compared with the
predictions of different theories. The quantities $\epsilon_1$, $\epsilon_2$,
$\epsilon_3$ and
$\epsilon_b$ are defined in ref.~\refnote{\cite{24}} in one-to-one correspondence
with the set of observables $m_W/m_Z$, $\Gamma_l$,
$A^{FB}_l$    and $R_b$. The four epsilons are defined without need of specifying
$m_t$ and $m_H$. In the SM, for all observables at the $Z$ pole, the whole
dependence on $m_t$ and $m_H$ arising from one-loop diagrams only enters through
the epsilons. The same is true for any extension of the SM such that all possible
deviations  only occur through vacuum polarization diagrams and/or the
$Z\rightarrow b\bar b$ vertex. 

	The epsilons represent an efficient parametrization of the small deviations from
what is solidly established, in a way that is unaffected by our relative
ignorance of $m_t$ and $m_H$. The variables $S, T, U$ depend on $m_t$ and $m_H$
because they are defined as deviations from the complete SM prediction for
specified $m_t$ and $m_H$. Instead the epsilons are defined with respect to a
reference approximation, which does not depend on $m_t$ and $m_H$.  In fact the
epsilons are defined in such a way that they are exactly zero in the SM in the
limit of neglecting all pure weak loop-corrections (i.e. when only the
predictions from the tree level SM plus pure QED and pure QCD corrections are
taken into account). This very simple version of improved Born approximation is a
good first approximation  according to the data. Values of the epsilons in the SM
are given in table 6
\refnote{\cite{23},\cite{24}}.
\begin{table} Table 6: Values of the epsilons in the SM as functions of $m_t$ and
$m_H$ as obtained from recent versions\refnote{\cite{23}} of ZFITTER  and TOPAZ0.
These values (in
$10^{-3}$ units) are obtained for
$\alpha_s(m_Z)$ = 0.118, 
$\alpha(m_Z)$ = 1/128.87, but the theoretical predictions are essentially
independent of
$\alpha_s(m_Z)$ and $\alpha(m_Z)$
 \refnote{\cite{24}}.
\begin{center}
\begin{tabular}{|c|l|l|l|l|l|l|l|l|l|c|}
\hline
$m_t$ & \multicolumn{3}{|c|}{$\epsilon_1$}&\multicolumn{3}{|c|}{$\epsilon_2$}
&\multicolumn{3}{|c|}{$\epsilon_3$}&$\epsilon_b$\\ (GeV)& \multicolumn{3}{|c|}
{$m_H$ (GeV) =} &  \multicolumn{3}{|c|} {$m_H$ (GeV) =} & \multicolumn{3}{|c|}
{$m_H$ (GeV) =} & All {$m_H$}\\ & 65 & 300 & 1000 & 65 & 300 & 1000 & 65 & 300 &
1000 &\\
\hline 150	&3.47&	2.76	& 1.61 &	$-$6.99 &	$-$6.61 &	$-$6.4 &	4.67	& 5.99 &	6.66 &
$-$4.45 \\ 160 &	4.34 &	3.59 &	2.38 &	$-$7.29 &	$-$6.9 &	$-$6.69 &	4.6 &	5.91 &
6.55 &	$-$5.28
\\
 170 &	5.25 &	4.46 &	3.21 &	$-$7.6 &	$-$7.2 &	$-$6.97 &	4.52 &	5.82 &	6.43 &
$-$6.13\\
 180 &	6.2 &	5.37 &	4.1 &	$-$7.93 &	$-$7.51 &	$-$7.24 &	4.42 &	5.72 &	6.34 &
$-$7.02\\
 190 &	7.2 &	6.33 &	5.07 &	$-$8.29 &	$-$7.81 &	$-$7.49 &	4.31 &	5.6 &	6.26 &
$-$7.95\\
 200 &	8.26 &	7.34 &	6.1 &	$-$8.65 &	$-$8.12 &	$-$7.75 &	4.19 &	5.49 &	6.19 &
$-$8.92\\
\hline
\end{tabular}
\end{center}
\end{table}

	By combining the value of $m_W/m_Z$ with the LEP results on the charged lepton
partial width and the forward--backward asymmetry, all given in table 1, and
following the definitions of ref.~\refnote{\cite{24}}, one obtains: 
\bea
			\epsilon_1 & = &\Delta\rho = (4.3 \pm 1.4)\times10^{-3}\nonumber
\\                                              
		 \epsilon_2 &= &(-6.9 \pm 3.4 ) \times 10^{-3}\nonumber
\\                                        
			\epsilon_3 &=& (3.0 \pm 1.8) \times
10^{-3}~.                                             
\label{18} 
\eea  Finally, by adding the value of $R_b$ listed in table 1 and using the
definition of $\epsilon_b$ given in ref.~\refnote{\cite{24}} one finds (note that
$\epsilon_b$ is defined through $R_b$ and the expression of
$R_b$ as a function of $\epsilon_b$ is practically independent of $\alpha_s$):
 \beq
		\epsilon_b = (-1.1\pm2.8) \times 10^{-3}~~~~~	(R_b~ {\rm from~table ~1})~. 
\label{19}
\eeq  This is the value that corresponds to the official average reported in
table 1 which I have criticized. Here in this epsilon analysis  we prefer to use
the ALEPH value for
$R_b$, ($R_b$ = 0.2161(14)), which leads to 
\beq
		\epsilon_b = (-5.7\pm3.4) \times 10^{-3}~~~~~	(\rm{R_b ~from ~ALEPH}) 
\label{20} 
\eeq
		
	To proceed further and include other measured observables in the analysis, we
need to make some dynamical assumptions. The minimum amount  of model dependence
is introduced by including other purely leptonic quantities at the $Z$ pole such
as
$A_{\tau_{pol}}$, $A_e$ (measured  from the angular dependence of the $\tau$
polarization) and $A_{LR}$ (measured by SLD). For this step, one is  simply
relying on lepton universality. Note that the choice of $A^{FB}_l$ as one of the
defining variables appears at present not particularly lucky, because the
corresponding determination of
$\sin^2\theta_{eff}$ markedly underfluctuates with respect to the average value
(see fig.~1). We then use the combined value of $\sin^2\theta_{eff}$ obtained
from the whole set of asymmetries measured at LEP and SLC, with the error
increased according to eq.~(\ref{8}) and the related discussion. At this stage
the best values of
$\epsilon_1$, $\epsilon_2$, $\epsilon_3$ and $\epsilon_b$ are modified according
to
\bea
			\epsilon_1 &=& \Delta\rho = (4.7\pm1.3) \times 10^{-3}\nonumber
\\                                              
		 \epsilon_2 &=& (-7.8\pm3.3 ) \times 10^{-3}\nonumber
\\                                       
			\epsilon_3 &=& (4.8\pm1.4)\times 10^{-3}\nonumber    \\
   \epsilon_b &=& (-5.7\pm3.4)\times 10^{-3}
\label{21} 
\eea
	In fig.~3  we report the 1$\sigma$ ellipse in the $\epsilon_1$--$\epsilon_3$
plane that correspond to this set of input data.
			                                                       
\begin{figure}
\hglue2.0cm
\epsfig{figure=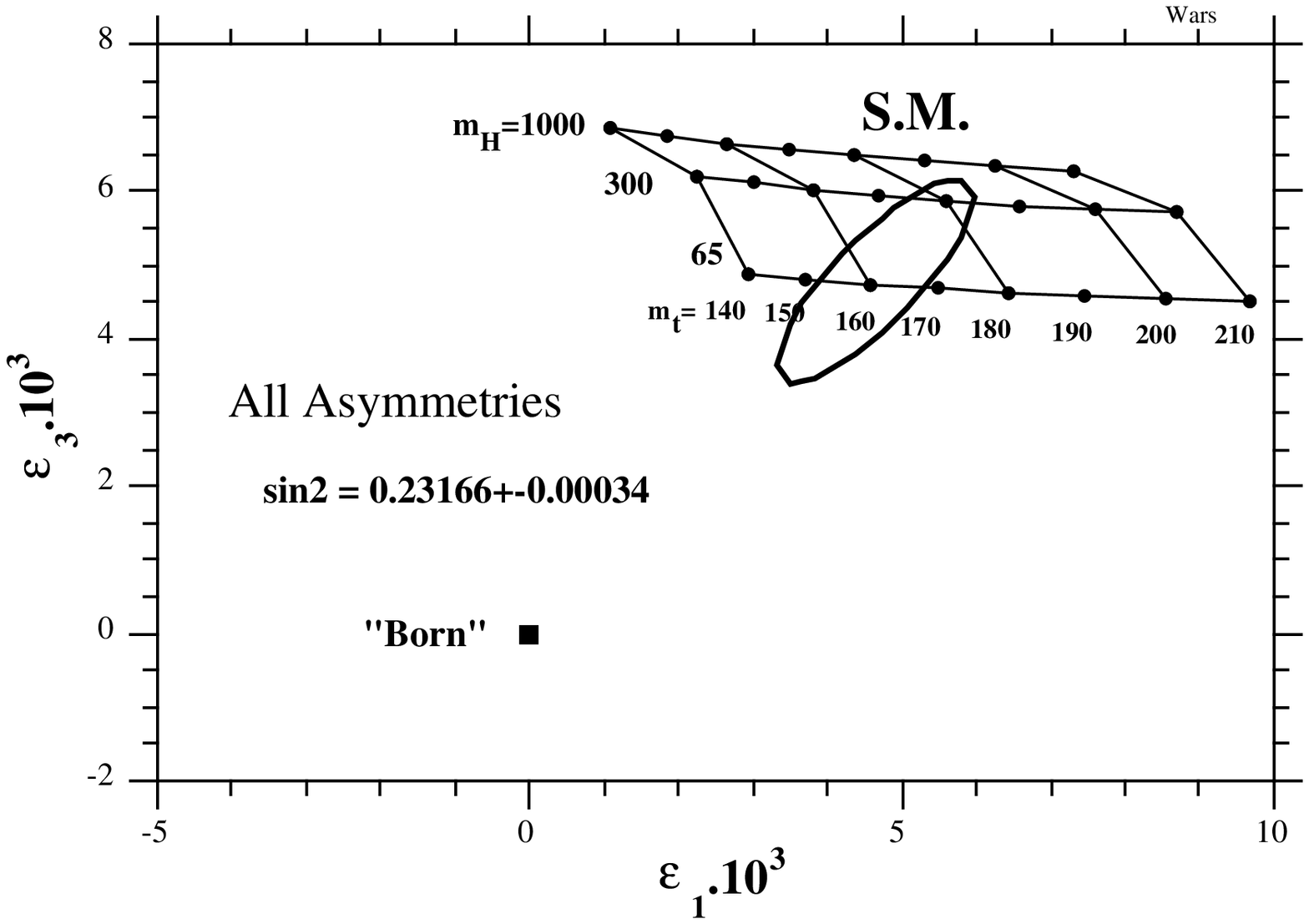,width=11.0cm}
\begin{center} Figure 3 \end{center}
\end{figure}

	All observables measured on the $Z$ peak at LEP can be included in the analysis,
provided that we assume that all deviations from the SM are only contained in
vacuum polarization diagrams (without demanding a truncation of the $q^2$
dependence of the corresponding functions) and/or the $Z\rightarrow b\bar b$ 
vertex. From a global fit of the data on $m_W/m_Z$,  $\Gamma_T$,  $R_h$,
$\sigma_h$,  $R_b$ and
$\sin^2\theta_{eff}$ (for LEP data, we have taken the correlation matrix for
$\Gamma_T$,  $R_h$ and
$\sigma_h$ given by the LEP experiments\refnote{\cite{ew}}, while we have
considered the additional information on $R_b$ and $\sin^2\theta_{eff}$  as
independent), we obtain:
\bea
			\epsilon_1 &=& \Delta\rho = (4.7\pm1.3) \times
10^{-3}\nonumber\\                                              
		 \epsilon_2 &=& (-7.8\pm3.3 ) \times
10^{-3}\nonumber\\                                        
			\epsilon_3 &=& (4.7\pm1.4) \times 10^{-3}\nonumber    \\
   \epsilon_b &=& (-4.8\pm3.2) \times
10^{-3}                                           
\label{22} 
\eea  The comparison of theory and experiment in the planes
$\epsilon_1$--$\epsilon_3$ and
$\epsilon_b$--$\epsilon_3$ is shown in figs.~4 and 5, respectively. Note that
adding the hadronic quantities hardly makes a difference in the
$\epsilon_1$--$\epsilon_3$ plot in comparison with fig.~3 which only included the
leptonic variables. In other words the inclusive hadronic quantities do not show
any peculiarity. A number of interesting features are clearly visible from this
plot. First, the good agreement with the SM and the evidence for weak
corrections, measured by the distance of the data from the improved Born
approximation point (based on tree level SM plus pure QED or QCD corrections).
Second, we see the preference for light Higgs or,  equivalently, the tendency for
$\epsilon_3$ to be rather on the low side (both features are now somewhat less
pronounced than they used to be). Finally, if the Higgs is light the preferred
value of $m_t$ is somewhat lower than the Tevatron result (which in this analysis
is not included in the input data). The data ellipse in the
$\epsilon_b$--$\epsilon_3$ plane is consistent with the SM and the CDF/D0 value of
$m_t$. This is because we have taken the ALEPH value for $R_b$.  For comparison,
we also show in figs.~6 and 7 the same plots as in figs.4 and 5, but for the
official average values of $R_b$ and
$\sin^2\theta_{eff}$ as reported in table 1. The main difference is the obvious
displacement of
$\epsilon_b$ and the smaller errors in the $\epsilon_1$--$\epsilon_3$ plot.
Finally, the status of
$\epsilon_2$ is presented in fig.~8. The agreement is very good. $\epsilon_2$ is
sensitive to $m_W$ and a more precise test will only be possible when the
measurement of $m_W$ will be much improved at LEP2 and the Tevatron.
 
\begin{figure}
\hglue2.0cm
\epsfig{figure=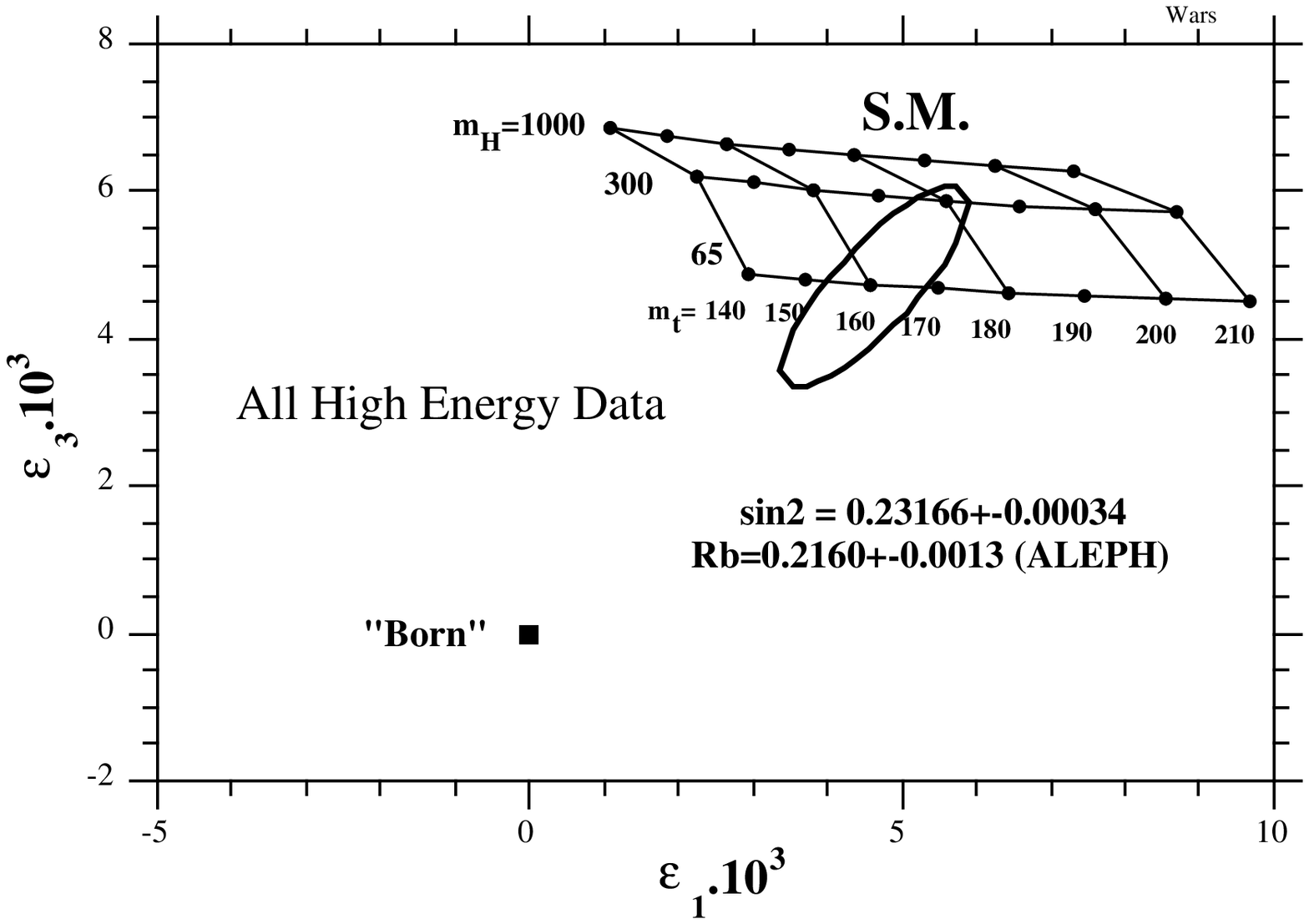,width=11.0cm}
\begin{center} Figure 4 \end{center}
\end{figure}

\begin{figure}
\hglue2.0cm
\epsfig{figure=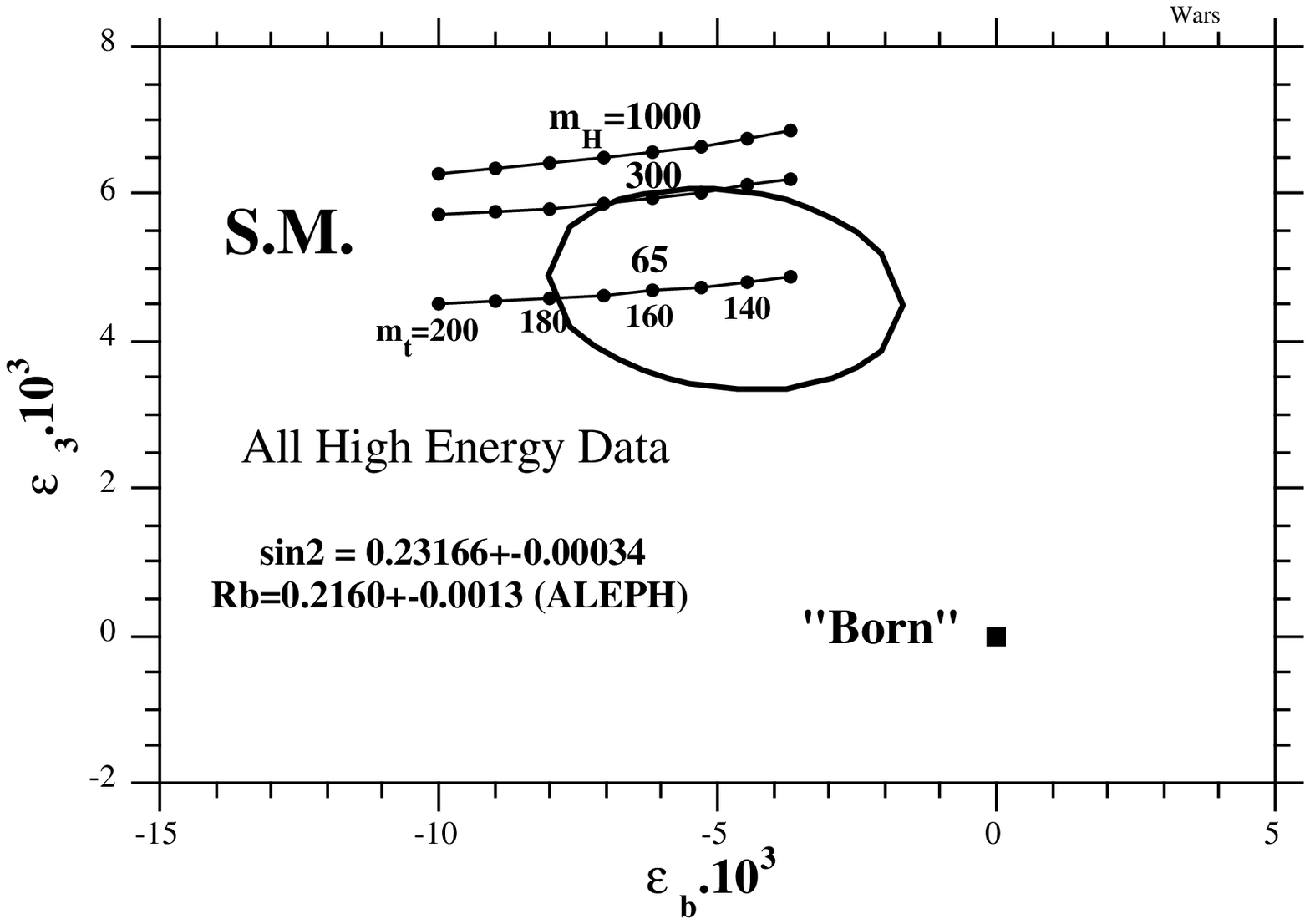,width=11.0cm}
\begin{center} Figure 5 \end{center}
\end{figure}

	 To include in our analysis lower-energy observables as well, a stronger
hypothesis needs to be made:  vacuum polarization diagrams are allowed to vary
from the SM  only in their constant and first derivative terms in a $q^2$
expansion\refnote{\cite{27}--\cite{29}}, a likely picture, e.g. in technicolor
theories\refnote{\cite{30}--\cite{32}}. In such a case, one can, for example, add
to the analysis the ratio
$R_\nu$ of neutral to charged current processes in deep inelastic neutrino
scattering on nuclei\refnote{\cite{33}}, the ``weak charge" $Q_W$  measured in
atomic parity violation experiments on Cs \refnote{\cite{34}},  and the
measurement of $g_V/g_A$ from
$\nu_\mu e$ scattering\refnote{\cite{35}}. In this way one obtains  the global fit
($R_b$ from ALEPH,
$\sin^2\theta_{eff}$  with enlarged error as in eq.~(\ref{8})): 
\bea
			\epsilon_1 &=& \Delta\rho = (4.3\pm1.2)
\times10^{-3}\nonumber\\                                              
		 \epsilon_2 &=& (-8.0\pm3.3 ) \times 10^{-3}\nonumber
\\                                        
			\epsilon_3 &=& (4.4\pm1.3) \times 10^{-3}\nonumber    \\
   \epsilon_b &=& (-4.6\pm3.2) \times 10^{-3}~.
\label{23} 
\eea

With the progress of LEP, the low-energy data, while important as a check that no
deviations from the expected
$q^2$ dependence arise, play a lesser role in the global fit. Note that the
present ambiguity on the value of $\delta\alpha^{-1}(m_Z) =\pm0.09$
\refnote{\cite{21}} corresponds to an uncertainty on $\epsilon_3$ (the other
epsilons are not much affected) given by
$\Delta\epsilon_3~10^3 =\pm0.6$ \refnote{\cite{24}}. Thus the theoretical error is
still comfortably less than the experimental error. 

	To conclude this section I would like to add some comments. As is clearly
indicated in figs.~3 to 8 there is by now solid evidence for departures from the
``improved Born approximation" where all the epsilons vanish. In other words a
strong evidence for the pure weak radiative corrections has been obtained, and
LEP/SLC are now measuring the various components of these radiative corrections.
For example, some authors
\refnote{\cite{39}} have studied the sensitivity of the data to a particularly
interesting subset of the weak radiative corrections, i.e. the purely bosonic
part. These terms arise from the virtual exchange of gauge bosons and Higgses.
The result is that indeed the measurements are sufficiently precise to require
the presence of these contributions in order to fit the data.
\begin{figure}
\hglue2.0cm
\epsfig{figure=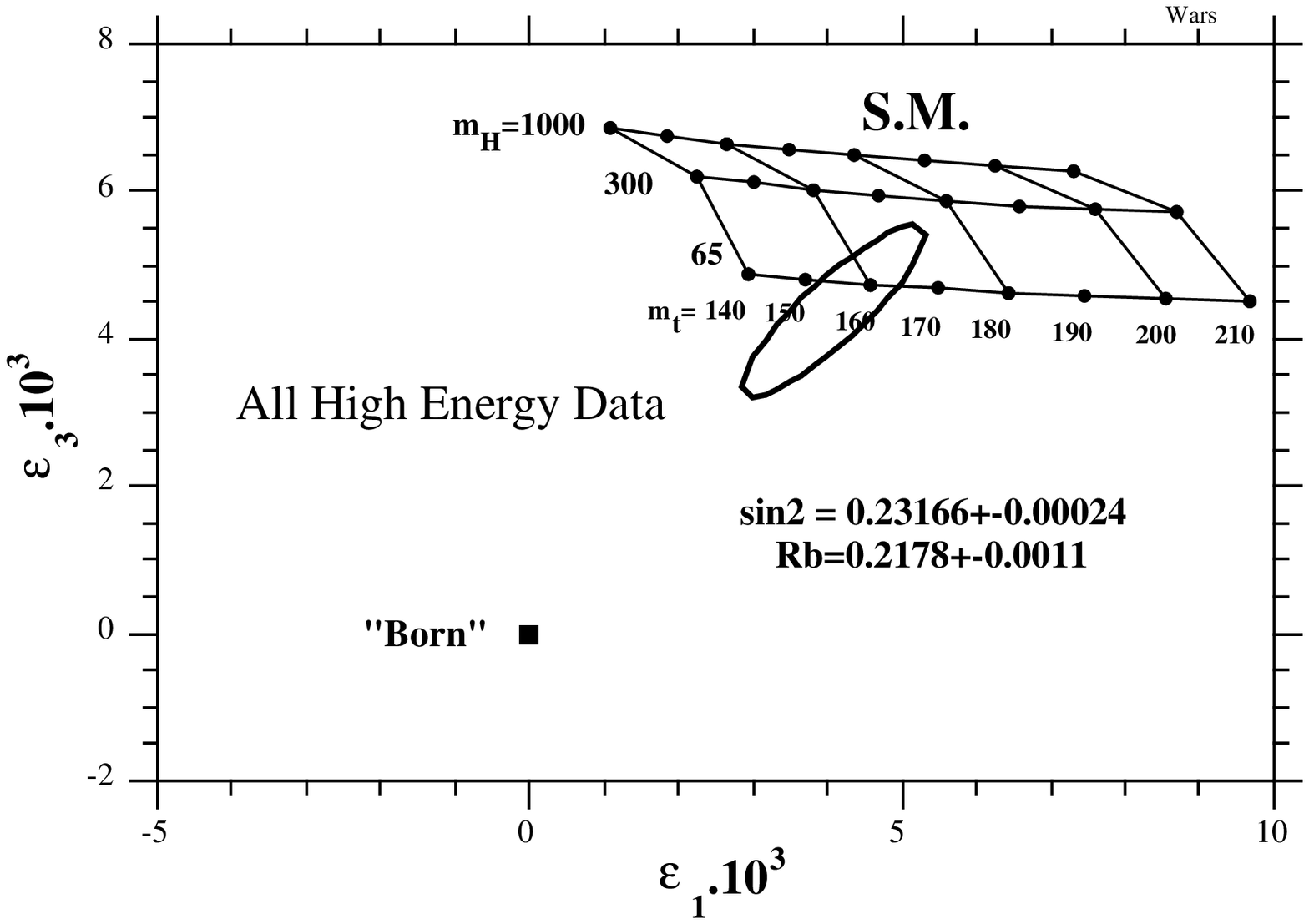,width=11.0cm}
\begin{center} Figure 6 \end{center}
\end{figure}

\begin{figure}
\hglue2.0cm
\epsfig{figure=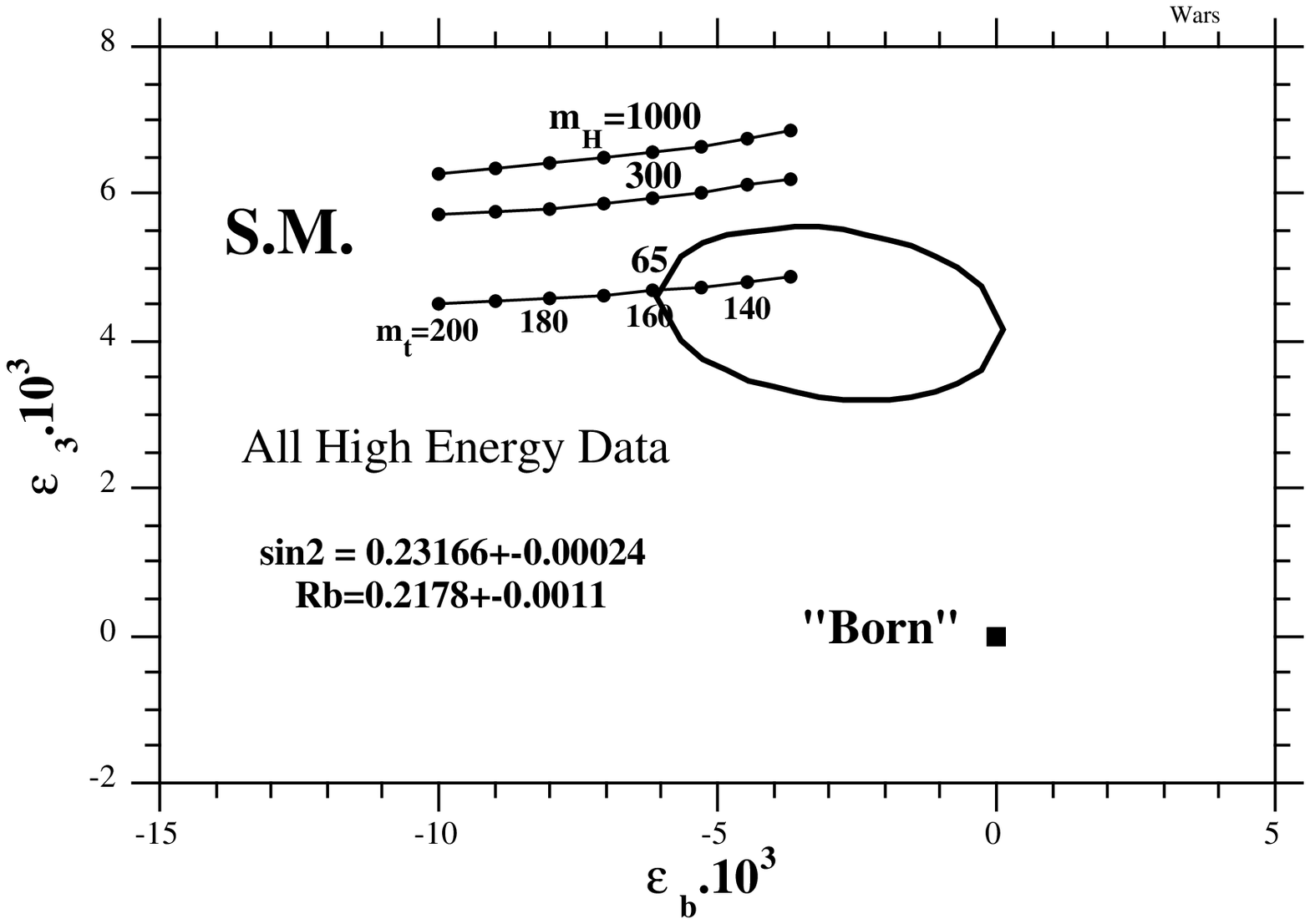,width=10.5cm}
\begin{center} Figure 7 \end{center}
\end{figure}

\begin{figure}
\hglue2.0cm
\epsfig{figure=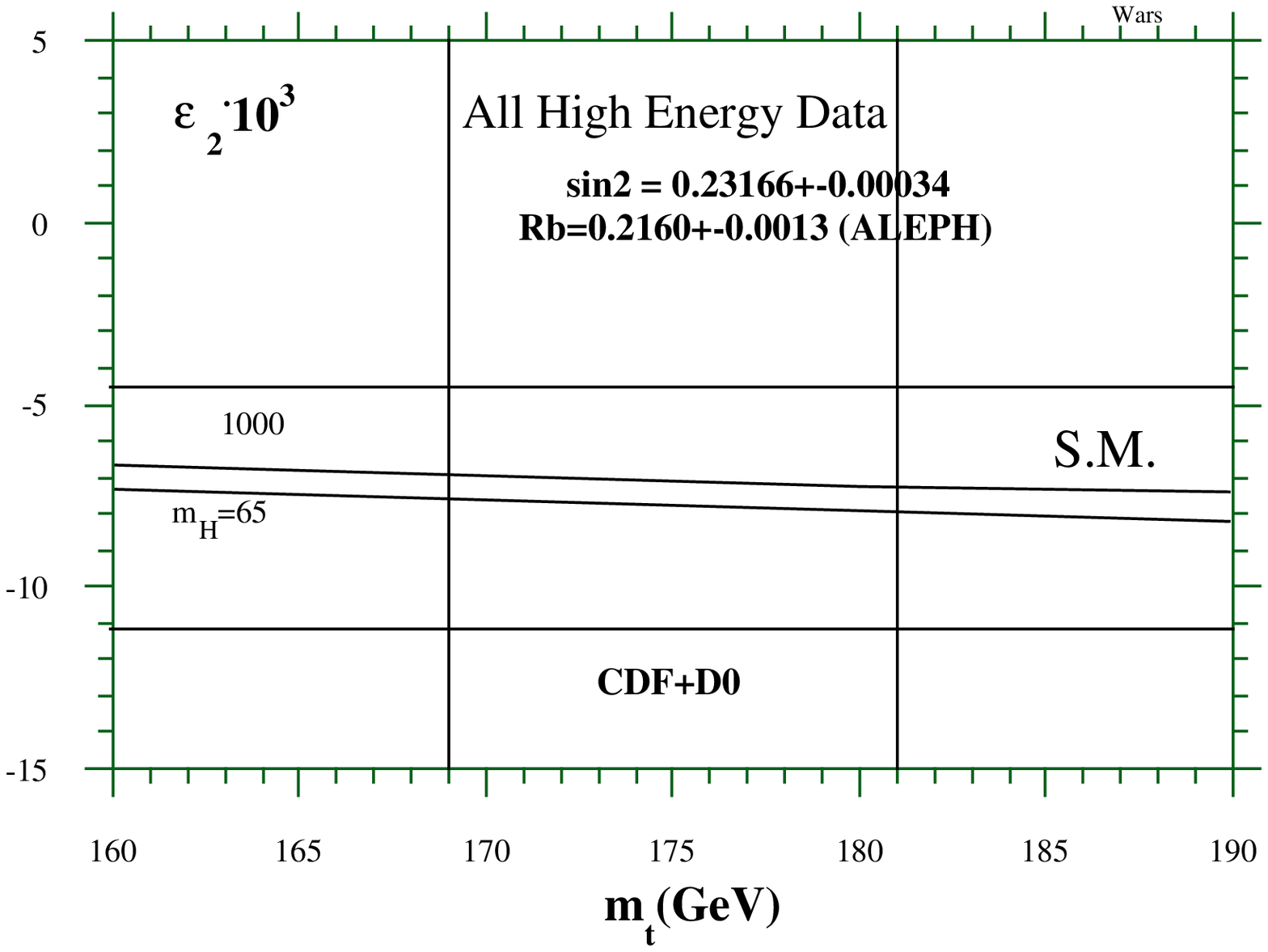,width=10.5cm}
\begin{center} Figure 8 \end{center}
\end{figure}

\section{ CONCEPTUAL PROBLEMS WITH THE STANDARD MODEL}

Given the striking success of the SM, why are we not satisfied with that theory?
Why not just find the Higgs particle, for completeness, and declare that particle
physics is closed? The main reason is that there are strong conceptual
indications for physics beyond the SM. 

	It is considered highly implausible that the origin of the electroweak symmetry
breaking can be explained by the standard Higgs mechanism, without accompanying
new phenomena. New physics should be manifest at energies in the TeV domain. This
conclusion follows fron an extrapolation of the SM at very high energies. The
computed behaviour of the $SU(3)\otimes SU(2)\otimes U(1)$ couplings with energy
clearly points towards the unification of the electroweak and strong forces
(Grand Unified Theories: GUTs) at scales of energy
$M_{GUT}\sim  10^{14}$--$10^{16}$~GeV, which are close to the scale of quantum
gravity,
$M_{Pl}\sim 10^{19}$~GeV
\refnote{\cite{qqi}}.  One can also imagine  a unified theory of all interactions
also including gravity (at present superstrings
\refnote{\cite{rri}} provide the best attempt at such a theory). Thus GUTs and the
realm of quantum gravity set a very distant energy horizon that modern particle
theory can no longer ignore. Can the SM without new physics be valid up to such
large energies? This appears unlikely because the structure of the SM could not
naturally explain the relative smallness of the weak scale of mass, set by the
Higgs mechanism at $m\sim 1/\sqrt{G_F}\sim  250$~GeV, $G_F$ being the Fermi
coupling constant. The weak scale $m$ is $\sim 10^{17}$ times smaller than 
$M_{Pl}$. Even if the weak scale is set near 250~ GeV at the classical level,
quantum fluctuations would naturally shift  it up to where new physics starts to
apply, in particular up to  $M_{Pl}$ if there was no new physics up to gravity.
This so-called hierarchy problem\refnote{\cite{ssi}} is related to the presence
of fundamental scalar fields in the theory with quadratic mass divergences and no
protective extra symmetry at $m=0$. For fermions, first, the divergences are
logaritmic and, second, at $m=0$ an additional symmetry, i.e. the chiral 
symmetry, is restored. Here, when talking of divergences we are not worried of
actual infinities. The theory is renormalizable and finite once the dependence on
the cut off is absorbed in a redefinition of masses and couplings. Rather the
hierarchy problem is one of naturalness. If we consider the cut off as a
manifestation of new physics that will modify the theory at large energy scales,
then it is relevant to look at the dependence of physical quantities on the cut
off and to demand that no unexplained enormously accurate cancellation arise. 

	According to the above argument the observed value of $m\sim 250$~GeV is
indicative of the existence of new physics nearby. There are two main
possibilities. Either there exist fundamental scalar Higgses, but the theory is
stabilized by supersymmetry, the boson--fermion symmetry that would downgrade the
degree of divergence from quadratic to logarithmic. For approximate supersymmetry
the cut off is replaced by the splitting between the normal particles and their
supersymmetric partners. Then naturalness demands that this splitting (times the
size of the weak gauge coupling) is of the order of the weak scale of mass, i.e.
the separation within supermultiplets should be of the order of no more than a
few TeV. In this case the masses of most supersymmetric partners of the known
particles, a very large menagerie of states, would fall, at least in part, in the
discovery reach of the LHC. There are consistent, fully formulated field theories
constructed on the basis of this idea, the simplest one being the
MSSM\refnote{\cite{43}}. Note that all normal observed states are those whose
masses are forbidden in the limit of exact
$SU(2)\otimes U(1)$. Instead, for all SUSY partners the masses are allowed in that
limit. Thus when supersymmetry is broken in the TeV range, but $SU(2)\otimes
U(1)$ is intact only spartners take mass while all normal particles remain
massless. Only at the lower weak scale the masses of ordinary particles are
generated. Thus a simple criterion exists to understand the difference between
particles and sparticles.

The other main avenue is compositeness of some sort. The Higgs boson is not
elementary but either a bound state of fermions or a condensate, due to a new
strong force, much stronger than the usual strong interactions, responsible for
the attraction. A plethora of new ``hadrons", bound by the new strong force,
would  exist in the LHC range. A serious problem for this idea is that nobody so
far has been  able to build up a realistic model along these lines, which could
eventually be explained by a lack of ingenuity on the theorists side. The most
appealing examples are technicolor theories\refnote{\cite{30},\cite{31}}. These
models where inspired by the breaking of chiral symmetry in massless QCD induced
by quark condensates. In the case of the electroweak breaking new heavy
techniquarks must be introduced and the scale analogous to $\Lambda_{QCD}$ must
be about three orders of magnitude larger. The presence of such a large force
relatively nearby has a strong tendency to clash with the results of the
electroweak precision tests\refnote{\cite{32}}. Another interesting idea is to
replace the Higgs by a
$t\bar t$ condensate\refnote{\cite{uui}}. The Yukawa coupling of the Higgs to the
$t\bar t$ pair becomes a four-fermion 
$\bar tt\bar tt$  coupling with the corresponding strength. The strong force is
in this case provided by the large top mass. At first sight this idea looks
great:  no fundamental scalars, no new states. But, looking closely, the
advantages are largely illusory. First, in the SM the required value of $m_t$ is
too large: $m_t\geq 220$~GeV or so. Also a tremendous fine-tuning is required,
because $m_t$ would naturally be of the order of $M_{GUT}$ or $M_{Pl}$ if no new
physics is present (the hierarchy problem in a different form!). Supersymmetry
could come to the rescue in this case also. In a minimal SUSY version the
required value of the top mass is lowered\refnote{\cite{vvi}},  $m_t\sim 195
\sin{\beta}$~ GeV. But the resulting theory is physically indistinguishable from
the MSSM with small
$\tan{\beta}$, at least at low energies\refnote{\cite{wwi}}. This is because a
strongly coupled Higgs looks the same as a $t\bar t$ pair.

	The hierarchy problem is certainly not the only conceptual problem of the SM.
There are many more: the proliferation of parameters, the mysterious pattern of
fermion masses and so on. But while most of these problems can be postponed to
the final theory that will take over at very large energies, of order $M_{GUT}$ or
$M_{Pl}$, the hierarchy problem arises from the instability of the low-energy
theory and requires a solution at relatively low energies. A supersymmetric
extension of the SM provides a way out that is well defined, computable and that
preserves all virtues of the SM. The necessary SUSY breaking can be introduced
through soft terms that do not spoil the stability of scalar masses. Precisely
those terms arise from supergravity when it is spontaneoulsly broken in a hidden
sector\refnote{\cite{yyi}}. But alternative mechanisms of SUSY breaking are also
being considered\refnote{\cite{gauge}}. As we shall now discuss, there are also
experimental and phenomenological hints that point in this direction.

	At present the most important phenomenological evidence in favour of
supersymmetry is obtained from the unification of couplings in GUTs. Precise LEP
data on $\alpha_s(m_Z)$ and
$\sin^2{\theta_W}$ confirm what was already known with less accuracy: standard
one-scale GUTs fail in predicting
$\sin^2{\theta_W}$ given
$\alpha_s(m_Z)$ (and $\alpha(m_Z)$), while SUSY GUTs\refnote{\cite{zzi}} are in
agreement with the present, very precise, experimental results. According to the
recent analysis of ref.~\refnote{\cite{aaii}}, if one starts from the known
values of
$\sin^2{\theta_W}$ and $\alpha(m_Z)$, one finds for $\alpha_s(m_Z)$ the results:
\bea
		\alpha_s(m_Z) &=& 0.073\pm 0.002 ~~~~~      	(\rm{Standard~ GUTS})\nonumber \\	
		\alpha_s(m_Z) &=& 0.129(+0.010,-0.008)~~~~~  (\rm{SUSY~ GUTS})
\label{24a}
\eea to be compared with the world average experimental value $\alpha_s(m_Z)$ =
0.118(5).

	A very elegant feature of the GUT-extended supersymmetric version of the SM is
that the occurrence of the
$SU(2)\otimes U(1)$ electroweak symmetry breaking is naturally and automatically
generated by the large mass of the top quark\refnote{\cite{bbii}}. Assuming that
all scalar masses are the same at the GUT scale, the effect of the large Yukawa
coupling of the top quark in the renormalization group evolution  down to the
weak energy scale, drives one of the Higgs squared masses negative (that Higgs
which is coupled to the up-type quarks). The masses of sleptons and of the Higgs
coupled to the down-type quark are much less modified, while the squark masses
are increased due to the strongly interacting gluino exchange diagrams. The
negative value of the squared mass corresponds to the onsetting of the
electroweak symmetry breaking. That the correct mass for the weak bosons is
obtained as a result of the breaking implies constraints on the model, more
stringent if no fine-tuning is allowed to a given level of accuracy. Various
fine-tuning criteria have been analysed in the
literature\refnote{\cite{ccii},\cite{ddii}}. Typically no more than a factor 10
fine tuning is allowed. With this assumption and realistic values of $m_t$ one
obtains the bounds shown in fig.~9\refnote{\cite{eeii}}. These upper bounds give
a quantitative specification of the constraints implied by a natural solution of
the hierarchy problem in the context of the GUT-extended MSSM. They look very
promising for LEP2 (but the bounds scale with the inverse square root of the
fine-tuning factor...).

\begin{figure}
\hglue0.75cm
\epsfig{figure=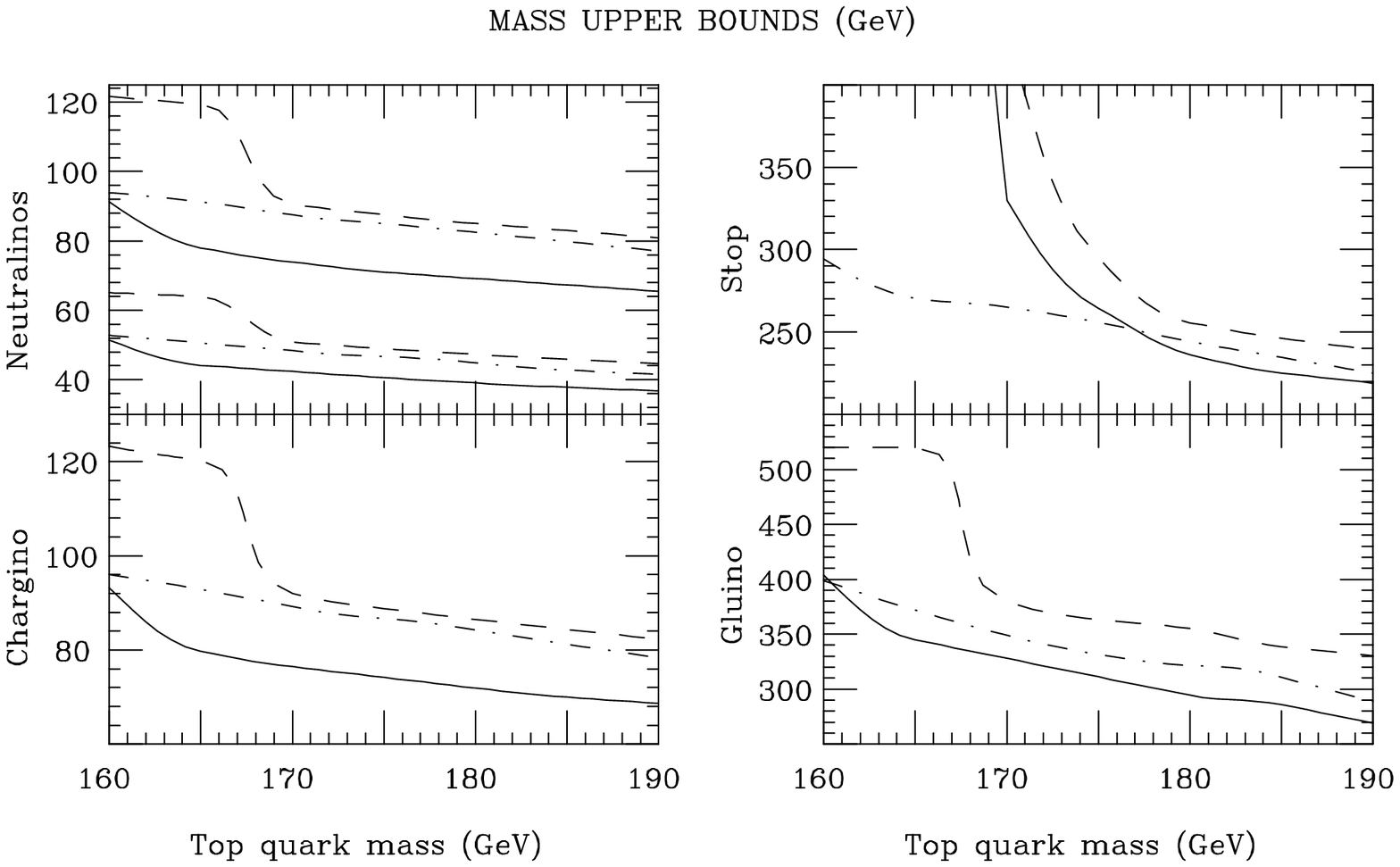,width=14cm}
\begin{center} Figure 9 \end{center}
\footnotesize{Upper bounds on gluino, lightest and next-to-lightest neutralino,
and lightest chargino and stop masses based on the requirement of no fine tuning
larger than 10\%.  The solid (dashed) lines refer to the minimal supersymmetric
standard model with universal boundary conditions at $M_{GUT}$ for the soft
supersymmetry-breaking terms, without (with) the inclusion of the one-loop
effective potential.  The dot-dashed lines show the mass upper limits, for
non-universal boundary conditions at $M_{GUT}$, without the includion of the
one-loop effective potential.}
\end{figure}
	Many of the simpler GUTs predict the unification at $M_{GUT}$ of the $b$ and
$\tau$ Yukawa couplings, or, equivalently, that for the running masses
$m_b(M_{GUT}) = m_\tau(M_{GUT})$
\refnote{\cite{ffii}}. The observed difference of the $b$ and $\tau$ masses
arises from the evolution due to the different interactions of quarks and
leptons. Many authors studied the combined constraints from coupling unification
and $b$ and $\tau$ Yukawa unification\refnote{\cite{ggii}}. The result is that there
are a small
$\tan{\beta}$ solution (typically in the range $\tan{\beta} = $0.5--3) and a large
$\tan{\beta}$ solution (with $\tan{\beta} = $40--60). However the large
$\tan{\beta}$  solution is somewhat disfavoured by a natural implementation of the
electroweak symmetry breaking, according to the mechanism discussed above. In
fact at large values of $\tan{\beta}\geq m_t/m_b$, the dominance of the top over
the bottom Yukawa coupling, which is an important ingredient for that mechanism,
is erased or even inverted. A closer look at the small $\tan{\beta}$ solution
shows that the top mass is close to its fixed-point solution $m_t\sim195
\sin{\beta}$~GeV so that
$m_t\sim175$~GeV corresponds to $\tan{\beta}\sim2$. Correspondingly the mass of
the lightest Higgs is relatively small\refnote{\cite{hhii}}, as discussed in
sect.~9, which is good for LEP2.

	In the MSSM the lightest neutralino is stable and provides a very good cold dark
matter candidate. It is interesting that if the constraint $\Omega$ = 1, which
corresponds to the critical density for closure of the Universe, is added to the
previous ones, consistency can still be achieved in a sizeable domain of the
parameter space\refnote{\cite{ggii}--\cite{jjii}}.

	In conclusion, gauge coupling unification, natural $SU(2)\otimes U(1)$
electroweak symmetry breaking, $b$ and $\tau$ Yukawa  unification and a plausible
amount of dark matter all fit together for realistic values of $m_t$,
$\alpha_s(m_Z)$ and $m_b$. SUSY GUTs, with a single step of symmetry breaking from
$m_{GUT}$ down to $m_W$,  appear to work well.

\section{PRECISION ELECTROWEAK TESTS AND THE SEARCH FOR NEW PHYSICS}
 
	We now concentrate on some well-known extensions of the SM, which not only are
particularly important per se but also are interesting in that they clearly
demonstrate the constraining power of the present level of precision tests. 

\subsection{Minimal Supersymmetric Standard Model} The MSSM\refnote{\cite{43}} is
a completely specified, consistent and computable theory. There are too many
parameters to attempt a direct fit of the data to the most general framework. So
one can consider two significant limiting cases: the ``heavy" and the ``light"
MSSM.

	The ``heavy" limit corresponds to all sparticles being sufficiently massive,
still within the limits of a natural explanation of the weak scale of mass. In
this limit a very important result holds\refnote{\cite{58}}: for what concerns
the precision electroweak tests, the MSSM predictions tend to reproduce the
results of the SM with a light Higgs, say
$m_H\lappeq$ 100 GeV.

	In the ``light" MSSM option, some of the superpartners have a relatively small
mass, close to their experimental lower bounds. In this case the pattern of
radiative corrections may sizeably deviate from that of the SM. The most
interesting effects occur in vacuum polarization amplitudes and/or the
$Z\rightarrow b\bar b$  vertex and therefore are particularly suitable for a
description in terms of the epsilons (because in such a case, as explained in
ref.~\refnote{\cite{24}}, the predictions can be compared with the experimental
determination of the epsilons from the whole set of LEP data). They are:
\begin{itemize}
\item[	i)] a threshold effect in the $Z$ wave function
renormalization\refnote{\cite{58}}, mostly due to the vector coupling of
charginos and (off-diagonal) neutralinos to the $Z$ itself. Defining the vacuum
polarization functions by
$\Pi_{\mu\nu}(q^2)=-ig_{\mu\nu}[A(0)+q^2 F(q^2)]+q_\mu q_\nu$ terms, this is a
positive contribution to $\epsilon_5=m^2_Z  F'_{ZZ} (m^2_Z)$,  the prime denoting
a derivative with respect to
$q^2$ (i.e. a contribution to a higher derivative term not included in the usual
$S, T, U$ formalism). The
$\epsilon_5$ correction shifts $\epsilon_1$, $\epsilon_2$ and $\epsilon_3$ by
$-\epsilon_5$, $-c^2\epsilon_5$ and  $-c^2\epsilon_5$ respectively, where
$c^2=\cos^2{\theta_W}$, so that all of them are reduced by a comparable amount.
Correspondingly all the Z widths are reduced without affecting the asymmetries.
This effect falls down particularly fast when the lightest chargino mass
increases from a value close to $m_Z$/2. Now that we know,  from the LEP1.5 and
LEP2 runs, that the chargino mass is not so light, its possible impact is
drastically reduced.
 
\item[ii)] A positive contribution to $\epsilon_1$ from the virtual exchange of
the scalar top and bottom  superpartners\refnote{\cite{59}}, analogous to the
contribution of the top--bottom left-handed quark doublet. The needed isospin
splitting requires one of the two scalars (in the MSSM the stop) to be light.
From the value of $m_t$, not much space is left for this possibility. If the stop
is light then it must be mainly a right-handed stop.

\item[iii)] A negative contribution to $\epsilon_b$, due to the virtual exchange
of a charged Higgs\refnote{\cite{60}}. If one defines, as customary,
$\tan{\beta}=v_2/v_1$ ($v_1$ and $v_2$ being the vacuum expectation values of the
Higgs doublets giving masses to the down and up quarks, respectively), then, for
negligible bottom Yukawa coupling or
$\tan{\beta}\geq m_t/m_b$, this contribution is proportional to
$m^2_t$ /$\tan^2{\beta}$.

\item[iv)] A positive contribution to $\epsilon_b$ due to virtual chargino--stop
exchange\refnote{\cite{61}}, which in this case is proportional to
$m^2_t/\sin^2{\beta}$ and prefers small $\tan\beta$. This effect again requires
the chargino and the  stop to be light in order to be sizeable.

\item[v)] A positive contribution to $\epsilon_b$ due to virtual $h$ and $A$
exchange\refnote{\cite{62}}, provided that
$\tan\beta$ is so large that the Higgs couplings to the $b$ quarks are as large
or more than to the $t$ quark.
\end{itemize}
	If there is really an excess in $R_b$, it could be explained by either of the
last two mechanisms\refnote{\cite{pok}}. For small $\tan\beta$ there is a positive
contribution to $R_b$
\refnote{\cite{60}--\cite{73}} if charginos and stops are light and the charged
Higgs is heavy. Not to spoil the agreement for
$\epsilon_1 = \Delta\rho$, we need the right stop to be light, while the left
stop and the sbottom are kept heavy and near to one another, which is quite
possible. Alternatively, for $\tan\beta$ large, of order 30 to 60, if $h$ and
$A$, the two neutral Higgses that can be lighter than the $Z$, are particularly
light, then one also obtains\refnote{\cite{62}} a substantial positive
contribution to $R_b$. The large
$\tan\beta$ value is needed in order to have a large coupling to
$b\bar b$. However, such large values of $\tan\beta$ are somewhat unnatural. Also
in this case having light charginos and stop helps.

\subsection{Technicolor}
 It is well known that technicolor models\refnote{\cite{30}--\cite{32}} tend to
produce large and positive corrections to
$\epsilon_3$. From ref.~\refnote{\cite{32}} where the data on $\epsilon_3$ and
$\epsilon_1$ are compared with the predictions of a class of simple versions of
technicolor models, one realizes that the experimental errors on $\epsilon_3$ are
by now small enough that these models are hopelessly disfavoured with respect to
the SM. 

	More recently it has been  shown\refnote{\cite{76}} that the data on
$\epsilon_b$ also produce evidence against technicolor models. The same mechanism
that in extended technicolor generates the top quark mass also leads to large
corrections to the 
$Z\rightarrow b\bar b$  vertex that have the wrong sign. For example, in a simple
model with two technidoublets ($N_{TC}$ = 2), the SM prediction is decreased by
the amount\refnote{\cite{76},\cite{77}}: 
\beq
			\Delta\epsilon_b = - 28\times 10^{-3} \left |\frac{\xi}{\xi'}\left
(\frac{m_t}{174~{\rm GeV}}\right )
\right|                                         
\label{24b}
 \eeq
 where $\xi$ and $\xi'$ are Clebsch-like coefficients, expected to be of order 1.
The effect is even larger for larger $N_{TC}$. In a more sophisticated version of
the theory, the so-called ``walking" technicolor\refnote{\cite{77}}, where the
relevant coupling constants walk (i.e. they evolve slowly) instead of running,
the result is somewhat smaller\refnote{\cite{78}} but still gigantic. Later it
was shown\refnote{\cite{79}} that in order to avoid this bad prediction one could
endow the extended technicolor currents with a non-trivial behaviour under the
electroweak group. 

	In conclusion, it is difficult to really exclude technicolor because it is not a 
completely defined theory, and no realistic model could be built so far out of
this idea. Yet, it is interesting that the most direct realizations tend to
produce
$\Delta\epsilon_3 \gg 0$ and $\Delta\epsilon_b \gg 0$ which are both disfavoured
by experiment. 

\section{OUTLOOK ON THE SEARCH FOR NEW PHYSICS}

	As we have seen in the previous sections, the whole set of electroweak tests is
at present quite consistent with the SM. The pattern of observed pulls shown in
table 1 accurately matches what we expect from a normal distribution of
measurement errors. Even the few hints of new physics that so far existed have
now vanished: $R_c$ is back to normal and $R_b$ is closer to the SM prediction.
We no longer need new physics to explain $R_b$. Even the faint indication that
$\alpha_s(m_Z)$ would prefer an excess in $R_b$ has disappeared. Of course it is
not excluded that a small excess of $R_b$ is indeed real. For example the chances
of nearby SUSY have not really been hit. Actually, with the absence of chargino
signals at LEP1.5 and LEP2, which implies an increase of the lower bound on the
chargino mass,  the most plausible range for a possible effect on $R_b$ in the
MSSM is bounded within
$\sim1\sigma$ or $\sim1.5\sigma$ of the ALEPH result (or $R_b\leq 0.2175$--0.2180)
\refnote{\cite{pok}}. 

	What is the status of other possible signals of new physics? The ALEPH multijet
signal at LEP1.5\refnote{\cite{7}} awaits confirmation from LEP2 before one can
really get excited. So far no such convincing confirmation has been reported from
the first
$\sim10~{\rm pb}^{-1}$ of integrated luminosity collected at LEP2 at
$\sqrt{s}$  = 161 GeV. The ALEPH multijet signal\refnote{\cite{7}}, if real,
cannot be interpreted in the MSSM. But it could be a signal of some more
unconventional realization of supersymmetry (e.g. with very light
gluinos\refnote{\cite{74}} or, more likely, with $R$-parity breaking
\refnote{\cite{75}}). It is perhaps premature to speculate on these events: in a
few months we will know for sure if they are real or not, as soon as LEP2 will
collect enough luminosity.

	The CDF excess of jets at large transverse energy is not very convincing
either\refnote{\cite{cdfjet}}. It is presented as an excess with respect to the
QCD prediction. But the QCD prediction can be to some extent forced in the
direction of the data by modifying the parton densities, in particular the gluon
density. At the price of a somewhat unnatural shape of the gluon density one can
sizeably reduce the discrepancy without clashing with other
data\refnote{\cite{91}}. On the contrary this is not the case for the quark
densities, which are tightly constrained by deep inelastic scattering data in the
same $x$ range\refnote{\cite{92}}. Also the newly released D0 data do not show
any additional evidence for the effect\refnote{\cite{93}}. However, the D0
results are less accurate. Thus on the one hand one can say that D0 is compatible
with either QCD or CDF. On the other hand their data are flat so that, to explain
the absence of the signal,  one should imagine a cancellation between the effect
and the variation of systematics with $E_T$. It was pointed out in
refs.~\refnote{\cite{85},\cite{86}} that if the effect was real it could be
explained in terms of a new vector boson $Z'$ of mass around 1~TeV coupled mainly
to quarks rather than leptons. In the presence of simultaneous anomalies in
$R_b$, $R_c$  and the jet yield at large
$E_T$, it was attractive to present a unique explanation for all three effects.
Now if only the jet excess is what remains this solution has lost most of its
appeal. But in principle it is still possible to reduce the mixing of the $Z'$ to
the ordinary $Z$ in such a way that its effect is only pronounced for jets while
it remains invisible at LEP\refnote{\cite{man}}.

	It is representative of the present situation that perhaps the best hint for new
physics in the data is given by the single CDF event with $e^+e^-\gamma\gamma
E\llap{$/$}_T$  in the final state\refnote{\cite{cdfsel}}. Indeed this event is
remarkable and it is difficult to imagine a SM origin for it. It is true that it
is easier to imagine an experimental misinterpretation of the event (e.g. a fake
electron, two events in one or the like) than a SM process that generates it. But
it is a single event and even an extremely unlikely possibility can occur once. 
Several papers have already been devoted to this event\refnote{\cite{pap}}. In
SUSY models two main possibilities have been investigated. Both interpret the
event as a selectron pair production followed by decays $\tilde e\rightarrow
eN'$,$N'\rightarrow N\gamma$. The observed production rate and the kinematics 
demand a selectron around 100 GeV and large branching ratios. In the first
interpretation, within the MSSM, $N'$ and $N$ are neutralinos. In order to make
the indicated modes dominant one has to restrict to a very special domain of the
parameter space of the model. Neutralinos and charginos in the LEP2 range are
then favoured. The second interpretation is based on the newly revived
alternative approach in which SUSY breaking is mediated by ordinary gauge rather
than gravitational interactions\refnote{\cite{43},\cite{gauge}}. In the most
familiar approach of the MSSM, SUSY is broken in a hidden sector and the scale of
SUSY breaking is very large, of order
$\Lambda\sim\sqrt{G^{-1/2}_F M_P}$,  where
$M_P$ is the Planck mass. But since the hidden sector only communicates with the
visible sector through gravitational interactions the splitting of the SUSY
multiplets is much smaller, in the TeV energy domain, and the goldstino is
practically decoupled. In the alternative scenario the (not so much) hidden
sector is connected to the visible one by ordinary gauge interactions. As these
are much stronger than the gravitational interactions,
$\Lambda$ can be much smaller, as low as 10--100 TeV. It follows that the
goldstino is very light in these models (with mass of order or below 1 eV
typically) and is the lightest, stable SUSY particle, but its couplings are
observably large. Then, in the CDF event,  $N'$ is a neutralino and $N$ is the
goldstino. The signature of photons comes out more naturally in this SUSY
breaking pattern than in the MSSM. If the event is really due to selectron
production it would be a manifestation of nearby SUSY that could be confirmed at
LEP2. This is what we all wish. We shall see!

\section{THE LEP2 PROGRAMME}

The LEP2 programme started at the end of June'96. At first the energy was fixed at
161~GeV, which is the most favourable energy for the measurement of $m_W$ from the
cross-section for
$e^+e^- \rightarrow W^+W^-$ at threshold. Then gradually the energy will be
increased up to a maximum of about 193~GeV to be reached in mid '98. An average
integrated luminosity of about 150~pb$^{-1}$  per year is foreseen. LEP2 will run
until the end of 1999 at least, before the shutdown for the installation of the
LHC. The main goals of LEP2 are the search for the Higgs and for new particles,
the measurement of $m_W$ and the investigation of the triple gauge vertices $WWZ$
and $WW\gamma$.  A complete updated survey of the LEP2 physics is collected in
two volumes\refnote{\cite{lep2}}. 

	An important competitor of LEP2 is the Tevatron collider. By and around the year
2000 the Tevatron will have collected about 1~fb$^{-1}$ of integrated luminosity
at 1.8--2~TeV. The competition is especially on the search of new particles, but
also on
$m_W$ and the triple gauge vertices. For example, for supersymmetry while the
Tevatron is superior for gluinos and squarks,  LEP2 is strong on Higgses,
charginos, neutralinos and sleptons.

	Concerning the Higgs it is interesting to recall that the large value of $m_t$
has important implications on
$m_H$ both in the minimal SM \refnote{\cite{zziii}--\cite{bbiiii}} and in its
minimal supersymmetric extension\refnote{\cite{cciiii},\cite{ddiiii}}. I will now
discuss the restrictions on $m_H$ that follow from the CDF value of $m_t$.

	It is well known\refnote{\cite{zziii}-\cite{bbiiii}} that in the SM with only one
Higgs doublet a lower limit on
$m_H$ can be derived from the requirement of vacuum stability. The limit is a
function of $m_t$ and of the energy scale $\Lambda$ where the model breaks down
and new physics appears. Similarly an upper bound on $m_H$ (with mild dependence
on $m_t$) is obtained\refnote{\cite{eeiiii}} from the requirement that up to the
scale $\Lambda$ no Landau pole appears. The lower limit on
$m_H$ is particularly important in view of the search for the Higgs at LEP2.
Indeed the issue is whether one can reach the conclusion that if a Higgs is found
at LEP2, i.e. with $m_H \leq m_Z$, then the SM must break down at some scale
$\Lambda  >$ 1~TeV. 

	The possible instability of the Higgs potential $V[\phi]$ is generated by the
quantum loop corrections to the classical expression of $V[\phi]$. At large
$\phi$ the derivative $V'[\phi]$ could become negative and the potential would
become unbound from below. The one-loop corrections to $V[\phi]$ in the SM are
well known and change the dominant term at large $\phi$ according to $\lambda
\phi^4 \rightarrow (\lambda +
\gamma~{\rm log}~\phi^2/\Lambda^2)\phi^4$. The one-loop approximation is not
enough for our purposes, because it fails at large enough $\phi$, when
$\gamma~{\rm log}~\phi^2/\Lambda^2$ becomes of order 1. The renormalization group
improved version of the corrected potential leads to the replacement
$\lambda\phi^4 \rightarrow
\lambda(\Lambda)\phi'^4(\Lambda)$, where $\lambda(\Lambda)$ is the running
coupling and
$\phi'(\mu) = {\rm exp}\int^t \gamma(t')dt'\phi$, with $\gamma(t)$ being an
anomalous dimension function and $t = {\rm log}\Lambda/v$ ($v$ is the vacuum
expectation value
$v = (2\sqrt 2 G_F)^{-1/2}$). As a result, the positivity condition for the
potential amounts to the requirement that the running coupling $\lambda(\Lambda)$
never becomes negative. A more precise calculation, which also takes into account
the quadratic term in the potential, confirms that the requirements of positive
$\lambda(\Lambda)$ leads to the correct bound down to scales $\Lambda$ as low as
$\sim$~1~TeV. The running of
$\lambda(\Lambda)$ at one loop is given by: 
\begin{equation}
\frac{d\lambda}{dt} = \frac{3}{4\pi^2} [ \lambda^2 + 3\lambda h^2_t - 9h^4_t +
{\rm gauge~terms}]~,
\label{24c}
\end{equation} with the normalization such that at $t=0, \lambda = \lambda_0 =
m^2_H/2v^2$ and the top Yukawa coupling $h_t^0 = m_t/v$. We see that, for $m_H$
small and $m_t$ large,
$\lambda$ decreases with $t$ and can become negative.  If one requires that
$\lambda$ remains positive up to $\Lambda = 10^{15}$--$10^{19}$~GeV, then the
resulting bound on $m_H$ in the SM with only one Higgs doublet is given
by\refnote{\cite{aaiiii}}:
\begin{equation} m_H > 135 + 2.1 \left[ m_t - 174 \right] -
4.5~\frac{\alpha_s(m_Z) - 0.118}{0.006}~.
\label{25}
\end{equation}

Summarizing, we see from Eq.~(\ref{25}) that indeed for $m_t > 150$~GeV the
discovery of a Higgs particle at LEP2 would imply that the SM breaks down at a
scale
$\Lambda$ below $M_{GUT}$ or $M_{Pl}$, smaller for lighter Higgs.  Actually, for
$m_t \sim$ 174~GeV, only a small range of values for $m_H$ is allowed, $130 < m_H
<~\sim 200$~GeV, if the SM holds up to $\Lambda \sim M_{GUT}$ or $M_{Pl}$ (where
the upper limit is from avoiding the Landau pole\refnote{\cite{eeiiii}}). As is
well known\refnote{\cite{zziii}} the lower limit is not much relaxed, even if
strict vacuum stability is replaced by some sufficiently long metastability. Of
course, the limit is only valid in the SM with one doublet of Higgses. It is
enough to add a second doublet to avoid the lower limit. A particularly important
example of theory where the bound is violated is the MSSM, which we now discuss. 

	As is well known\refnote{\cite{43}}, in the MSSM there are two Higgs doublets,
which implies three neutral physical Higgs particles and a pair of charged
Higgses. The lightest neutral Higgs, called $h$, should be lighter than
$m_Z$ at tree-level approximation. However, radiative
corrections\refnote{\cite{ffiiii}} increase the $h$ mass by a term proportional to
$m^4_t$ and  logarithmically dependent on the stop mass. Once the radiative
corrections are taken into account the $h$ mass still remains rather small: for
$m_t$ = 174~GeV one finds the limit (for all values of $\tan\beta)~m_h < 130$~GeV
\refnote{\cite{ddiiii}}. Actually there are reasons to expect that $m_h$ is well
below the bound. In fact, if $h_t$ is large at the GUT scale, which is suggested
by the large observed value ot $m_t$ and by a natural onsetting of the
electroweak symmetry breaking induced by $m_t$, then at low energy a fixed point
is reached in the evolution of $m_t$. The fixed point corresponds to $m_t \sim
195 \sin\beta$~GeV (a good approximate relation for  $\tan\beta = v_{up}/v_{down}
< 10$). If the fixed-point situation is realized, then $m_h$ is considerably
below the bound, as shown in Ref.~56.

In conclusion, for $m_t \sim 174$~GeV, we have seen that, on the one hand, if a
Higgs is found at LEP the SM cannot be valid up to $M_{Pl}$. On the other hand,
if a Higgs is found at LEP, then the MSSM has good chances, because this model
would be excluded for $m_h > 130$~GeV.

	For the SM Higgs, which plays the role of a benchmark, also important for a more
general context, the LEP2 reach has been studied in detail. Accurate simulations
have shown\refnote{\cite{lep2}} that at LEP2 with 500~pb$^{-1}$ per experiment,
or with 150~pb$^{-1}$ if the four experiments are combined, one can reach the
5$\sigma$ discovery range given by $m_H
\leq 82, 95$~GeV for $\sqrt s  = 175, 192$~GeV respectively, and the 95\%
exclusion range
$m_H \leq 83, 98$~GeV. On the basis of these ranges we understand why a few GeV
make a lot of difference. With
$\sqrt s = 175$~GeV there would be practically no overlap with the LHC, which,
even in the most optimistic projections, cannot see the Higgs below  $m_H =
80$~GeV or so. With $\sqrt s$ = 185~GeV, there starts to be some overlap, but
only limited to $m_H
\leq 85$--90~GeV, which is still a very difficult, time-consuming and debatable
range for the LHC. With
$\sqrt  s$ = 195~GeV there is already a quite reasonable overlap, up to $m_H
\leq$ 95--100~GeV.  The issue is not only that of avoiding a gap between LEP2 and
the LHC, but also of providing an essential independent and complementary channel
to study the new particle in a range of mass that is certainly rather marginal
for the LHC. 

	In the MSSM a more complicated discussion is needed because there are several
Higgses and the parameter space is multidimensional. Also, through the radiative
corrections, the Higgs masses at fixed values of all MSSM parameters sensitively
depend on the top quark mass. For decreasing top quark masses the upper bound on
the light Higgs mass decreases. We note that the discovery range for LEP2 can be
specified in terms of the light Higgs mass with little model dependence. On the
contrary the same analysis for the LHC depends very much on the detailed
quantitative pattern of the decay branching ratios. The usual plots that are seen
in the experimental discussions are based on some typical choice of parameters,
which is to some extent indicative.

	In Ref.~\refnote{\cite{lep2}}, the analysis for the MSSM is presented in great
detail, as this case is rather complicated and was not deeply studied previously.
With the typical choice of parameters, in the sense specified above, the domains
of the
$\tan\beta, m_A$ plane which are most difficult for the LHC are a ``hole" at
moderate values of $\tan\beta$ and $m_A$ (say  $\tan\beta < 10, m_A
=100$--200~GeV) and a ``strip" at small $\tan\beta$ and large
$m_A$ (typically $\tan\beta$ = 1--3 and $m_A > $ 300~GeV). If $m_t$ is not too
small, these difficult regions can probably be covered at the LHC, but only with
very large integrated luminosities $L= 3\times 10^5$~pb$^{-1}$. LEP2 potentially
can reduce the ``hole" and completely cover the ``strip", especially for
$m_t$ rather small. But while for $\sqrt s = 175$~GeV this is only true for
rather extreme values of $m_t$ and the squark mixing, at $\sqrt s = 192$~GeV only
the central values are required (always with 150~pb$^{-1}$ of integrated
luminosity and the four experiments combined). Thus, as in the case of the SM,
$\sqrt s = 192$~GeV is needed for a reasonable overlap, while less than that
appears risky. 

	We now consider the search for supersymmetry. For charginos the discovery range
at LEP2 is only limited by the beam energy for practically all values of the
parameters. Thus every increase of the beam energy is directly translated into
the upper limit in chargino mass for discovery or exclusion. For the Tevatron the
discovery range is much more dependent on the position in parameter space. For
some limited regions of this space, with 1~fb$^{-1}$ of integrated luminosity,
the discovery range for charginos at the Tevatron goes well beyond
$m_\chi =~$90--100~GeV, i.e. the boundary of LEP2, while in most of the parameter
space one would not be able to go that far and only LEP2, with sufficient energy,
would find the chargino. 

	The stop is probably the lightest squark. For a light stop the most likely decay
modes are  $\tilde t
\rightarrow b\chi^+$ if kinematically allowed, otherwise  
$\tilde t \rightarrow c \chi$. A comparative study of these modes at LEP2 and at
the Tevatron is presented in Ref.~\refnote{\cite{lep2}}. The result is that in
either case at LEP2 the discovery range is up to about $(E_{beam}-10)$~GeV. At
the Tevatron there is some difference between the two possible decay modes and
some dependence on the position in the 
$\tilde t$--$\chi$ or the $\tilde t$--$\chi^+$ planes, but it is true that very
soon, at the end of the present run, with 100~pb$^{-1}$, a large region of the
potential LEP2 discovery range will be excluded (in particular for the 
$\tilde t \rightarrow c \chi$ mode). Some limited regions will require more
luminosity at the Tevatron and could be accessible to LEP2.

	While on the stop the chances are better at the Tevatron than at LEP2 the
converse is true for the sleptons. Here the Tevatron can only compete for a
particularly favourable pattern of branching ratios. Finally, for neutralinos
there is only a small region of the parameter space where these particles would
be the first spartners to be discovered. The discovery ranges are very much
parameter-dependent both at the Tevatron and at LEP2. For these reasons no
detailed quantitative comparison still exists, although the channel
$e^+e^-\rightarrow \chi \chi'$ has been extensively studied at the LEP2
workshop\refnote{\cite{lep2}}.

The measurement of $m_W$ will be done at LEP2 from the cross-section at threshold
and from direct reconstruction of the jet--jet final state in $W$ decay. At
present
$m_W$ is known with an error of $\pm 150$~MeV from the direct measurement (see
table~1). From the fit to all electroweak data one finds $m_W = 80352 \pm 34$~MeV
(see eq.~(\ref10)), in agreement with the direct measurement. As a consequence
the goal for LEP2 is to measure $m_W$ with an accuracy $\delta m_W \leq \pm
(30-40)$~MeV, in order to provide an additional significant check of the theory. 

For the threshold method\refnote{\cite{lep2}} the minimum of the statistical
error is obtained for
$\sqrt s = 2m_W + 0.5$~GeV = 161 GeV, which in fact was the initial operating
energy of LEP2. The total error of this method is dominated by the statistics. If
each of the four experiments will eventually collect 50~pb$^{-1}$ of integrated
luminosity (10 already collected and the rest in a possible future comeback at
low energy) and the results are combined, then the statistical error will be
$\delta m_W =
\pm 95$~MeV and the total error $\delta m_W = \pm 108$~MeV. After
$\sim$~10~pb$^{-1}$ the present combined result is $m_W = (80.4\pm 0.2 \pm
0.1)$~GeV
\refnote{\cite{semi}}. Thus with realistic luminosity this method is not
sufficient by itself.

	In principle the direct reconstruction method can use the totally hadronic or
the semileptonic final states
$e^+e^- \rightarrow W^+W^- \rightarrow jjjj$ or $jjl\nu$. The total branching
ratio of the hadronic modes is 49\%, while that of the $\ell = e,\mu$
semileptonic channels is 28\%. The hadronic channel has more statistics but could
be severely affected by non-perturbative strong interaction effects: colour
recombination among the jets from different $W$'s and Bose correlations among
mesons in the final state from $WW$ overlap. Colour recombination is
perturbatively small. But gluons with $E < \Gamma_W$ are important and
non-perturbative effects could be relatively large, of the order of 10--100~MeV.
Similarly for Bose correlations. One is not in a position to really quantify the
associated uncertainties. Fortunately the direct reconstruction from the
semileptonic channels can, by itself, lead to a total error $\delta m_W = \pm
44$~MeV,  for the combined four experiments, each with 500~pb$^{-1}$ of
luminosity collected at $\sqrt s \geq 175$~GeV. Thus the goal of measuring $m_W$
with an accuracy below  $\delta m_W = \pm 50$~MeV can be fulfilled, and it is
possible to do better if one learns from the data how to limit the error from
colour recombination and Bose correlations.

	The study of triple gauge vertices is another major task of LEP2. The
capabilities of LEP2 in this domain are comparable to those of the LHC and go
well below the level of deviations from the tree-level couplings that in the SM
are expected from one-loop  radiative corrections. LEP2 can push down the
existing direct limits considerably. For given anomalous couplings the departures
from the SM are expected to increase with energy. For the energy and the
luminosity available at LEP2, given the accuracy of the SM established at LEP1,
it is however not very likely, to find signals of new physics in the triple gauge
vertices.

It is a pleasure for me to thank Tom Ferbel for his kind invitation and warm
hospitality in St.Croix.  It is with great sadness that I recall the memory of
George~Michail, a student at this School, a fine young man who died shortly
afterwards in a tragic accident.


\begin{numbibliography}
\bibitem{blo} A. Blondel, Proceedings of ICHEP '96, Warsaw, 1996;\\ M.
Pepe-Altarelli, Proceedings of the Symposium on Radiative Corrections, Cracow,
1996. 

\bibitem{ew} The LEP Electroweak Working Group, LEPEWWG/96-02.

\bibitem{sta} G. Altarelli, Proceedings of the Int. Symposium on Lepton and
Photon Interactions, Stanford, 1989.

 \bibitem{car} F. Caravaglios, hep-ph/9610416.

\bibitem{tip} P. Tipton, Proceedings of ICHEP '96, Warsaw, 1996.

\bibitem{10} G. Altarelli, R. Kleiss and C. Verzegnassi (eds.), Z  Physics at
LEPÊ1 (CERN 89-08, Geneva, 1989), Vols. 1--3. 

\bibitem{11}	Precision Calculations for the Z Resonance, eds. D. Bardin, W.
Hollik and  G. Passarino, Report CERN 95-03 (1995).

\bibitem{piet}  B. Pietrzyk, Proceedings of the Symposium on Radiative
Corrections, Cracow, 1996.
 
\bibitem{12}	F. Jegerlehner, {\it Z. Phys.} {\bf C32} (1986) 195.

\bibitem{13}	B.W. Lynn, G. Penso and C. Verzegnassi, {\it Phys. Rev.} {\bf D35}
(1987) 42.

\bibitem{14}	H. Burkhardt et al.,  {\it Z. Phys.} {\bf C43} (1989) 497.

\bibitem{15}	F. Jegerlehner, {\it Progr. Part. Nucl. Phys.} {\bf 27} (1991) 32.

\bibitem{16}	M.L. Swartz, Preprint SLAC-PUB-6710, 1994.

\bibitem{17}	M.L. Swartz, {\it Phys. Rev.} {\bf D53} (1996) 5268.

\bibitem{18}	A.D. Martin and D. Zeppenfeld, {\it Phys. Lett.} {\bf B345}(1995)
558.

\bibitem{19} 	R.D. Nevzorov and A.V. Novikov, Preprint FTUV/94-27, 1994, Preprint
MAD/PH/855, 1994.

\bibitem{20}	H. Burkhardt and B. Pietrzyk,  Preprint LAPP-EXP-95.05, 1995.

\bibitem{21}	S. Eidelman and F. Jegerlehner,  {\it Z. Phys.} {\bf C67} (1995) 585.

\bibitem{22} D. Haidt in ``Precision tests of the Standard Electroweak Model",
ed. P.~Langacker, World Scientific, Singapore, 1993.

\bibitem{wil} S. Willenbrock,Proceedings of ICHEP '96, Warsaw, 1996.

\bibitem{sigtop} G. Altarelli, M. Diemoz, G. Martinelli and P. Nason, {\it Nucl.
Phys.} {\bf B308} (1988) 724; \\
 R.K. Ellis, {\it Phys. Lett.} {\bf 259B} (1991) 492; \\ E. Laenen, J. Smith and
W.L. van Neerven, {\it Nucl. Phys.} {\bf B369} (1992) 543, {\it Phys. Lett.} {\bf
321B} (1994) 254; \\
 E. Berger and H. Contopanagos, {\it Phys. Lett.} {\bf 361B} (1995) 115; hep-ph
9512212;\\  S. Catani, M. Mangano, P. Nason and L. Trentadue, hep-ph 9602208.
 
\bibitem{23} ZFITTER: D. Bardin et al., CERN-TH. 6443/92 and refs. therein;\\
   	TOPAZ0: G. Montagna et al., {\it Nucl. Phys.} {\bf B401} (1993) 3; {\it Comp.
Phys. Comm.}  {\bf 76} (1993) 328.
	BHM: G. Burgers et al., LEPTOP: V.A. Novikov et al., WOH, W. Hollik :  see 	ref.
\cite{11}.

\bibitem{24}	G. Altarelli, R. Barbieri and S. Jadach, {\it Nucl. Phys.} {\bf B369}
(1992)3;\\
	G. Altarelli, R. Barbieri and F. Caravaglios, {\it Nucl. Phys.} {\bf B405}
(1993) 3;
 {\it Phys. Lett.} {\bf B349} (1995) 145.

\bibitem{gur}  A. Gurtu, Preprint TIFR-EHEP/96/01.

\bibitem{sch} M. Schmelling, Proceedings of ICHEP '96, Warsaw, 1996.

\bibitem{har} D. Harris (CCFR Collaboration),Proceedings of ICHEP '96, Warsaw,
1996.

\bibitem{virc} M. Virchaux and A. Milsztajn, {\it Phys. Lett.} {\bf B274} (1992)
221.

\bibitem{fly} J. Flynn, Proceedings of ICHEP '96, Warsaw, 1996.

\bibitem{gri} B. Grinstein and I.Z. Rothstein, hep-ph/9605260.

\bibitem{dal} A.X. El-Khadra, Proceedings of ICHEP '92, Dallas, 1992.

\bibitem{mich} C. Michael, Proceedings of the Int. Symposium on Lepton and Photon
interactions, Beijing, 1995.

\bibitem{26}	M.E. Peskin and T. Takeuchi, {\it Phys. Rev. Lett.} {\bf 65} (1990)
964 and {\it Phys. Rev.} {\bf D46} 	(1991) 381.

\bibitem{27}	G. Altarelli and R. Barbieri, {\it Phys. Lett.} {\bf B253} (1990)
161;\\
	B.W. Lynn, M.E. Peskin and R.G. Stuart, SLAC-PUB-3725 (1985);  in Physics at 
LEP, Report CERN 86-02, Vol. I, p. 90.

\bibitem{28}	B. Holdom and J. Terning, {\it Phys. Lett.} {\bf B247} (1990) 88;\\
	D.C. Kennedy and P. Langacker, {\it Phys. Rev. Lett.} {\bf 65} (1990) 2967 and
preprint  UPR-0467T;\\
	B. Holdom, Fermilab 90/263-T (1990);\\
	A. Ali and G. Degrassi, DESY preprint DESY 91-035 (1991);\\
	E. Gates and J. Terning, {\it Phys. Rev. Lett.} {\bf 67} (1991) 1840;\\
	E. Ma and P. Roy, {\it Phys. Rev. Lett.} {\bf 68} (1992) 2879;\\
	G. Bhattacharyya, S. Banerjee and P. Roy, {\it Phys. Rev.} {\bf D45} (1992) 729.

\bibitem{29}	M. Golden and L. Randall, {\it Nucl. Phys.} {\bf B361} (1991) 3;\\
	M. Dugan and L. Randall, {\it Phys. Lett.} {\bf B264} (1991) 154;\\
	A. Dobado et al., {\it Phys. Lett.} {\bf B255} (1991) 405;\\
	J. Layssac, F.M. Renard and C. Verzegnassi, Preprint UCLA/93/TEP/16 (1993).

\bibitem{30}	S. Weinberg, {\it Phys. Rev.} {\bf D13} (1976) 974 and {\it Phys.
Rev.} {\bf D19} (1979) 1277;\\
	L. Susskind, {\it Phys. Rev.} {\bf D20} (1979) 2619;\\
	E. Farhi and L. Susskind, {\it Phys. Rep.} {\bf 74} (1981) 277.

\bibitem{31}	R. Casalbuoni et al., {\it Phys. Lett.} {\bf B258} (1991) 161;\\
	R.N. Cahn and M. Suzuki, LBL-30351 (1991);\\
	C. Roisnel and Tran N. Truong, {\it Phys. Lett.} {\bf B253} (1991) 439;\\
	T. Appelquist and G. Triantaphyllou, {\it Phys. Lett.} {\bf B278} (1992) 345;\\
	T. Appelquist, Proceedings of the Rencontres de la Vall\'ee d'Aoste, La Thuile, 
Italy, 1993.

\bibitem{32}	J. Ellis, G.L. Fogli and E. Lisi, {\it Phys. Lett.} {\bf B343} (1995)
282.

\bibitem{33}	CHARM Collaboration, J.V. Allaby et al., {\it Phys. Lett.} {\bf B177}
(1986) 446; {\it  Z. Phys.} 	{\bf C36} (1987) 611;\\
	CDHS Collaboration, H. Abramowicz et al., {\it Phys. Rev. Lett.} {\bf 57} (1986)
298;\\ 
	 A. Blondel et al., {\it Z. Phys.}  {\bf C45} (1990) 361;\\
	CCFR Collaboration, A. Bodek, Proceedings of the EPS Conference on High 	Energy
Physics, Marseille, France, 1993.

\bibitem{34}	M.C. Noecker et al., {\it Phys. Rev. Lett.} {\bf 61} (1988) 310;\\
	M. Bouchiat, Proceedings of the 12th International Atomic Physics 	Conference
(1990).

\bibitem{35}	CHARM II Collaboration, R. Berger, Proceedings of the EPS Conference
on 	High Energy Physics, Marseille, France, 1993.

\bibitem{39}	S. Dittmaier, D. Schildknecht and G. Weiglein, {\it Nucl. Phys.}
{\bf B465} (1996) 3.

\bibitem{qqi} See, for example, G.G. Ross, ``Grand Unified Theories", Benjamin,
New York, 1984;\\ R. Mohapatra, {\it Prog. Part. Nucl. Phys.} {\bf 26} (1991) 1.

\bibitem{rri} See, for example, M.B. Green, J.H. Schwarz and E. Witten,
``Superstring Theory",  University Press, Cambridge, 1987.

\bibitem{ssi} E. Gildener, {\it Phys. Rev.} {\bf D14} (1976) 1667;\\
 E. Gildener and S. Weinberg, {\it Phys. Rev.} 	{\bf D15} (1976) 3333.

\bibitem{43} H.P. Nilles, {\it Phys. Rep.} {\bf C110} (1984) 1;\\ H.E. Haber and
G.L. Kane, {\it Phys. Rep.} {\bf C117} (1985) 75;\\ R. Barbieri, {\it Riv. Nuovo
Cim.} {\bf 11} (1988) 1.

\bibitem{uui} For a review, see, for example, C.T. Hill, in ``Perspectives on
Higgs Physics", ed. G.~Kane, World Scientific, Singapore, 1993, and references
therein.

\bibitem{vvi} W.A. Bardeen, T.E. Clark and S.T. Love, {\it Phys. Lett.} {\bf
B237} (1990) 235;\\ M. Carena et al., {\it Nucl. Phys.} {\bf B369} (1992) 33.

\bibitem{wwi} A. Hasenfratz et al., UCSD/PTH 91-06(1991).

\bibitem{yyi} A. Chamseddine, R. Arnowitt and P. Nath, {\it Phys. Rev. Lett.}
{\bf 49} (1982) 970;\\ R. Barbieri, S. Ferrara and C. Savoy, {\it Phys. Lett.}
{\bf 110B} (1982) 343;\\ E. Cremmer et al., {\it Phys. Lett.} {\bf 116B} (1982)
215.

\bibitem{zzi} S. Dimopoulos, S. Raby and F. Wilczek, {\it Phys. Rev.} {\bf D24}
(1981) 1681;\\
	S. Dimopoulos and H. Georgi, {\it Nucl. Phys.} {\bf B193} (1981) 150;\\ L.E.
Ib\'a$\tilde {\rm n}$ez and G.G. Ross, {\it Phys. Lett.} {\bf 105B} (1981) 439.

\bibitem{aaii} P. Langacker, Proceedings of the Tennessee International Symposium
on  Radiative Corrections ed. B.F.L.~Ward, Gatlinburg, USA, 1994, p. 415.

\bibitem{bbii} 	L.E. Ib\'a$\tilde {\rm n}$ez and G.G. Ross, {\it Phys. Lett.}
 {\bf 110B} (1982) 215;\\
	L. Alvarez-Gaum\'e, M. Claudson and M.B. Wise, {\it Nucl. Phys.}  {\bf B207}
(1982) 96.

\bibitem{ccii} R. Barbieri and G.F. Giudice, {\it Nucl. Phys.} {\bf B306} (1988)
63;\\ G.G. Ross and R.G. Roberts, {\it Nucl. Phys.} {\bf B377} (1992) 571.

\bibitem{ddii} B. de Carlos and J.A. Casas, CERN-TH.7024/93.

\bibitem{eeii} G. Altarelli et al., CERN-TH/95-151.

\bibitem{ffii} M.S. Chanowitz, J. Ellis and M.K. Gaillard, {\it Nucl. Phys.} {\bf
B128} (1977) 506;\\ A.J. Buras et al., {\it Nucl. Phys.} {\bf B135} (1978) 66.

\bibitem{ggii} V. Barger, M.S. Berger and P. Ohmann, {\it Phys. Rev.} {\bf D47}
(1993) 1093;\\ P. Langacker and N. Polonsky, {\it Phys. Rev.} {\bf D49} (1994)
1454;\\ M. Carena, S. Pokorski and C. Wagner, {\it Nucl. Phys.} {\bf B406} (1993)
59;\\ G.L. Kane et al. {\it Phys. Rev.} {\bf D49} (1994) 6173, {\bf D50} (1994)
3498;\\ W. de Boer et al., Karlsruhe preprint IEKP-KA/94-05 (1994).

\bibitem{hhii} M. Carena et al., {\it Nucl. Phys.} {\bf B419} (1994) 213.

\bibitem{jjii}  R.G. Roberts and L. Roszkowski, {\it Phys. Lett.} {\bf B309}
(1993) 329.

\bibitem{58}	R. Barbieri, F. Caravaglios and M. Frigeni, {\it Phys. Lett.} {\bf
B279} (1992) 169.

\bibitem{59}	R. Barbieri and L. Maiani, {\it Nucl. Phys.} {\bf B224} (1983) 32;\\
	L. Alvarez-Gaum\'e, J. Polchinski and M. Wise, {\it Nucl. Phys.} {\bf B221}
(1983) 495.

\bibitem{60}	W. Hollik, {\it Mod. Phys. Lett.} {\bf A5} (1990) 1909.

\bibitem{61}	A. Djouadi et al., {\it Nucl. Phys.} {\bf B349} (1991) 48;\\
	M. Boulware and D. Finell, {\it Phys. Rev.} {\bf D44} (1991) 2054.  The sign
discrepancy  between these two papers now appears to be solved in favour of the
second one.

\bibitem{62}	D. Garcia and J. Sola, {\it Phys. Lett.} {\bf B354} (1995) 335.

\bibitem{pok} S. Pokorski, Proceedings of ICHEP '96, Warsaw, 1996.
 
\bibitem{63}	G. Altarelli, R. Barbieri and F. Caravaglios, {\it Phys. Lett.} {\bf
B314} (1993) 357.

\bibitem{64}	J.D. Wells, C. Kolda and G. L. Kane,  {\it Phys. Lett.} {\bf B338}
(1994) 219.
 
\bibitem{65}	X. Wang, J.L.Lopez and D.V. Nanopoulos, {\it Phys. Rev.} {\bf D52}
(1995) 4116.
  
\bibitem{66}	G. Kane, R. Stuart and J. Wells, {\it Phys. Lett.} {\bf B354} (1995)
350; see also hep-ph/9510372.
 
\bibitem{67}	P. Chankowski and S. Pokorski, hep-ph/9505308.

\bibitem{68} 	Y. Yamada, K. Hagiwara and S. Matsumoto, hep-ph/9512227.

\bibitem{69}	E. Ma and D. Ng, hep-ph/9508338.
 
\bibitem{70}	A. Brignole, F. Feruglio and F. Zwirner, hep-ph9601293.

\bibitem{71}	G.F. Giudice and A. Pomarol, CERN-TH/95-337.

\bibitem{72}	J. Ellis, J.L. Lopez and D.V. Nanopoulos, CERN-TH/95-314
(hep-ph/9512288).
 
\bibitem{73}	P. Chankowski and S. Pokorski, hep-ph9603310.

\bibitem{76}	R.S. Chivukula, S.B. Selipsky and E.H. Simmons, {\it Phys. Rev.
Lett.} {\bf 69} (1992) 575.
 
\bibitem{77}	B. Holdom, {\it Phys. Lett.} {\bf 105} (1985) 301;\\
	K. Yamawaki, M. Bando and K. Matumoto, {\it Phys. Rev. Lett.} {\bf 56} (1986)
1335;\\
	V.A. Miransky, {\it Nuov. Cim.} {\bf 90A} (1985);\\
	T. Appelquist, D. Karabali and L.C.R. Wijewardhana,  {\it Phys. Rev.} {\bf D35}
(1987) 389; 	149;\\
	T. Appelquist and L.C.R. Wijewardhana, {\it Phys. Rev.} {\bf D35} (1987) 774;
	{\it Phys. Rev.} {\bf D36} (1987) 568.

\bibitem{78}	R.S. Chivukula et al., Preprint BUHEP-93-11 (1993).

\bibitem{79}	R.S. Chivukula, E.H. Simmons and J. Terning, Preprint BUHEP-94-08
(1994).

\bibitem{7}	ALEPH Collaboration, D. Buskolic et al., CERN/PPE/96-52.

\bibitem{74}	G. Farrar, hep-ph/9512306.

\bibitem{75}	A.K. Grant et al., hep-ph/960139253.
	
\bibitem{cdfjet} CDF Collaboration, F. Abe et al., Preprint
Fermilab-Pub-96/020-E(1996).

\bibitem{91}	J. Huston et al., Michigan Preprint, MSU-HEP-50812;\\ 
	W.K. Tung, Proceedings of DIS'96, Rome, 1996.

\bibitem{92}	E.W.N. Glover et al., hep-ph/9603327.

\bibitem{93}	D0 Collaboration, Proceedings of the Rencontres de Moriond, Les Arcs,
1996.

\bibitem{85}	P. Chiappetta, J. Layssac, F.M. Renard and C.~Verzegnassi,
hep-ph/9601306.

\bibitem{86}	G. Altarelli, N. DiBartolomeo, F. Feruglio, R.~Gatto and M.~Mangano,
{\it Phys. Lett.} {\bf B375} (1996) 292.

\bibitem{cdfsel} CDF Collaboration ,S. Park, Proceedings of the Workshop on 
Proton-Antiproton Collider Physics, ed. R. Raja and J. Yoh, AIP Press, 1995.

\bibitem{man} M. Mangano et al., Proceedings of ICHEP '96, Warsaw, 1996.

\bibitem{pap} S. Ambrosanio et al., {\it Phys. Rev. Lett.} {\bf 76} (1996) 3494;\\
	S. Dimopoulos et al., {\it Phys. Rev. Lett.} 76 (1996) 3498;\\ S. Ambrosanio and
B. Mele, hep-ph-9609212.

\bibitem{gauge} M. Dine and A.E. Nelson, {\it Phys. Rev.} {\bf D48} (1993) 1277;\\
M. Dine, A.E. Nelson and Y. Shirman, {\it Phys. Rev.} {\bf D51} (1995) 1362;\\ M.
Dine, A.E. Nelson, Y. Nir and Y. Shirman, {\it Phys. Rev.} {\bf D53} (1996)2658;\\
T. Gherghetta, G. Jungman and E. Poppitz, hep-ph/9511317;\\ G. Dvali, G.F.
Giudice and A. Pomarol, hep-ph/9603238.

\bibitem{lep2} G. Altarelli, T. Sj\"ostrand and F.~Zwirner (eds.), ``Physics at
LEP2", Report CERN 95-03.

\bibitem{zziii} 	M. Sher, {\it Phys. Rep.} {\bf 179} (1989) 273;  {\it Phys.
Lett.} {\bf B317} (1993) 159.

\bibitem{aaiiii} G. Altarelli and G. Isidori, {\it Phys. Lett.} {\bf B337} (1994)
141.

\bibitem{bbiiii} J.A. Casas, J.R. Espinosa and M. Quiros,  {\it Phys. Lett.} {\bf
B342} (1995) 171.

\bibitem{cciiii} J.A. Casas et al., {\it Nucl. Phys.} {\bf B436} (1995) 3;  E{\bf
B439} (1995) 466.

\bibitem{ddiiii} M. Carena and C.E.M. Wagner, {\it Nucl. Phys.} {\bf B452} (1995)
45.

\bibitem{eeiiii} See, for example, M. Lindner,  {\it Z. Phys.} {\bf 31} (1986)
295.

\bibitem{ffiiii} H. Haber and R. Hempfling, {\it Phys. Rev. Lett.} {\bf 66} (1991)
1815;\\ J. Ellis, G. Ridolfi and F. Zwirner, {\it Phys. Lett.} {\bf B257} (1991)
83;\\ Y. Okado, M. Yamaguchi and T. Yanagida, {\it Progr. Theor. Phys. Lett.}
{\bf 85} (1991) 1;\\
	R. Barbieri, F. Caravaglios and M. Frigeni, {\it Phys. Lett.} {\bf B258} (1991)
167.
	For a 2-loop improvement, see also: \\ R. Hempfling and A.H. Hoang, {\it Phys.
Lett.} {\bf B331} (1994) 99.

\bibitem{semi} N. Watson, presented in a seminar at CERN on 8~Oct.~1996.

\end{numbibliography}
\end{document}